\documentclass[letterpaper,preprintnumbers,preprint,aps,accepted=2022-05-26]{quantumarticle}
\pdfoutput=1

\usepackage{amsfonts}
\usepackage{amssymb} 
\usepackage{amsmath} 
\usepackage{bbold}
\usepackage{hyperref}
\usepackage[utf8]{inputenc}
\usepackage[qm,braket]{qcircuit}
\usepackage[dvipsnames,usenames]{xcolor}
\usepackage[ruled,vlined]{algorithm2e}
\usepackage{graphicx}
\usepackage{setspace}
\usepackage{tikz}
\usepackage[normalem]{ulem}

\usetikzlibrary{trees,positioning,fit,arrows,decorations.pathreplacing, calc, shapes.geometric}

\title{Quantum Machine Learning with SQUID}
\date{May 26, 2022}

\author{Alessandro Roggero}
\affiliation{Institute for Nuclear Theory, University of Washington, Seattle, WA 98195, USA}
\affiliation{InQubator for Quantum Simulation (IQuS), Department of Physics, University of Washington, Seattle, WA 98195, USA}

\author{Jakub Filipek}
\affiliation{Paul G. Allen School of Computer Science \& Engineering, University of Washington, Seattle, WA 98195, USA}

\author{Shih-Chieh Hsu}
\affiliation{Department of Physics, University of Washington, Seattle 98195, USA}

\author{Nathan Wiebe}
\affiliation{University of Toronto, Department of Computer Science, Toronto, ON M5G 1V7, Canada}
\affiliation{Pacific Northwest National Laboratory, Richland, WA 99352, USA}
\affiliation{Department of Physics, University of Washington, Seattle 98195, USA}
\begin{document}

\preprint{INT-PUB-21-010,IQuS@UW-21-006}

\maketitle

\begin{abstract}
In this work we present the Scaled QUantum IDentifier (SQUID), an open-source framework for exploring hybrid Quantum-Classical algorithms for classification problems. The classical infrastructure is based on PyTorch and we provide a standardized design to implement a variety of quantum models with the capability of back-propagation for efficient training. We present the structure of our framework and provide examples of using SQUID in a standard binary classification problem from the popular MNIST dataset. In particular we highlight the implications for scalability for gradient based optimization of quantum models on the choice of output for variational quantum models.
\end{abstract}

\section{Introduction}

Quantum Machine Learning (QML) is a rapidly growing, emerging field, with a diverse set of ideas and applications.
While there are significant differences in applications as to where Machine Learning and where Quantum Computing are applied, quantum-enhanced machine learning has become one of the dominant subfields~\cite{Dunjko2020,Wiebe2020,Guan_2021,Cerezo_2021}.
The main benefits of such algorithms are potential quantum speed-ups~\cite{Rebentrost2014,Biamonte2017,Huang_2021,Huang_2021B,Liu_2021}, and the potential of recognizing statistical patterns hard to learn with purely classical schemes~\cite{Schuld_2019b}. Recent work on QML has started to tackle the problems of trainability~\cite{McClean_2018,Beer_2020,marrero2021entanglement,Pesah2021} and the generalization error~\cite{Banchi2021,Du2022} of quantum models.

However, machine learning algorithms on near-term quantum devices face an issue of constrained resources.
While there exist encodings that efficiently use qubits, they still do not allow to load datasets such as MNIST in quantum memory, while additionally introducing overhead when encoding and decoding information between classical and quantum devices.
To counter that issue researchers have used two approaches.

One was to use synthetic or very small datasets that could be learned efficiently (as in eg.~\cite{Havl_ek_2019}).
This however makes any benchmarks artificial, and comparison to their classical counter-parts hard.
As potential performance benefits are the main driver of the quantum-enhanced machine learning, we believe that ease of comparison to classical machine learning should be one of the priorities in the field.

Second approach is to classically pre-process data, so it can fit in the limited space defining the quantum model (as in eg.~\cite{peters2021machine}).
While this approach allows for direct comparison to classical performance on the same data, it also requires to factor in what impact pre-processing had on performance of both algorithms.
It requires scientists to carefully prepare experiments to not give unfair advantages to either quantum or classical algorithms.
Lastly, since no two studies will use the same pre-processing there is additional overhead when comparing two different quantum-enhanced approaches, or performing meta-analysis of the field.

To combat these issues, we propose a standardized approach of designing hybrid (quantum and classical) models.
Similarly to how TensorFlow~\cite{tensorflow2015-whitepaper} and PyTorch~\cite{PyTorch} changed classical machine learning field and increased reproducibility of efforts, we propose Scaled QUantum IDentifier (SQUID)~\cite{squid}, which is an extensible framework which can incorporate quantum models.
As it is based on top of PyTorch, it has most of the benefits of a mature framework when it comes to purely classical architectures.
For quantum models, we provide a standardized model design where user has to implement forward- and back-propagation functions.

By doing so, the pre-processing algorithms can be standard across applications and approaches, making them more directly comparable.
It also reduces overhead on new researchers, as it significantly reduces the amount of coding required for an experiment.
Such mix of both worlds also resembles quantum-inspired algorithms~\cite{Dunjko2020}, which also benefit from above points.

The article is organized in following manner.
In Section~\ref{sec:squid} we outline the framework design and describe the relevant internal details.
In Section~\ref{sec:results} we show an example application of the model using the MNIST dataset, and study the impact of including information from single vs. all available output qubits. We describe the use of the SQUID Helpers package and possible future extensions in Sec.~\ref{sec:ext}. Finally, in Section.~\ref{sec:conclusions} we provide a summary and perspective.


\section{The squid framework}
\label{sec:squid}
Our main goal when designing SQUID is to propose a framework within which both classical and quantum machine learning can work in concert to solve a classification problem.
Properly utilizing classical computing, when possible, is of great importance because quantum and classical models for data will often have different advantages and disadvantages.
From an architectural perspective, the key innovation that SQUID allows is for classical neural networks to be globally trained in conjunction with a quantum neural network to build optimal encoders and decoders for the classical inputs or quantum outputs from the hybrid neural network.
This ability allows us to, in effect, learn a feature map that not only allows us to represent large quantum datasets in near term devices but also allows us to incorporate classical information that may be known a priori into the quantum model.

Before proceeding with the detailed description of the model, it is important for work like this that hybridizes between quantum and classical models to discuss the correspondence between the quantum and classical machine learning models that we implicitly assume.
In all these cases we assume that in general our training dataset contains both classical vector and quantum states and the following form
\begin{equation}
S_{\rm train} = \{(v_{\rm class}[j],\ket{v_{\rm quant}[j]}), j\in 0,\ldots,N_{\rm train} \}
\end{equation}
Here we assume a classical bit-encoding, meaning that we assume that each $v_{\rm class}[j] \in \mathbb{R}^D$.
This is the typical setting in machine learning, however, it is also possible to envision that the actual training vectors are distributions over $D$ symbols and the classical values $v_{\rm class}$ are given by the probabilities of drawing each symbol from the distribution.
We further implicitly assume that the quantum data $\ket{v_{\rm quant}[j]}$ is provided using an amplitude encoding.  By this we mean that the values of the training vectors are stored in the amplitudes of the $\ket{v_{\rm quant}[j]}$ state.
We make this choice because it is the most general setting that we can assume as it also subsumes the case where the quantum training vectors correspond to distinct quantum bit strings (otherwise known as a bit-encoding).

The SQUID framework was designed with extensibility and simplicity as its core principles.
It generally follows design principles set by PyTorch~\cite{PyTorch}.
Currently there are many competing Quantum SDKs~\cite{smith2016practical, Qiskit, Cirq}, most of which include Python interfaces.
Hence a successful QML package should allow, a simple extensible solution, which can be adjusted to any specific SDK.
SQUID enables that by providing general classes, similar to \texttt{nn.Module} in PyTorch, one each for Quantum and Classical Models.
These satisfy minimal requirements of functions used by the backend \texttt{MainModel} to properly propagate the gradient through combinations of the models.

\subsection{Framework Design}
Main component of SQUID is the \texttt{MainModel}, which itself accepts three smaller models.
The first and last models are currently enforced to be classical, while the middle can be either quantum or classical. In case all three models are classical, \texttt{MainModel} is equivalent to PyTorch's \texttt{nn.Sequential} with three sub-components.

\begin{figure}
	\centering
	\includegraphics[width=0.475\textwidth]{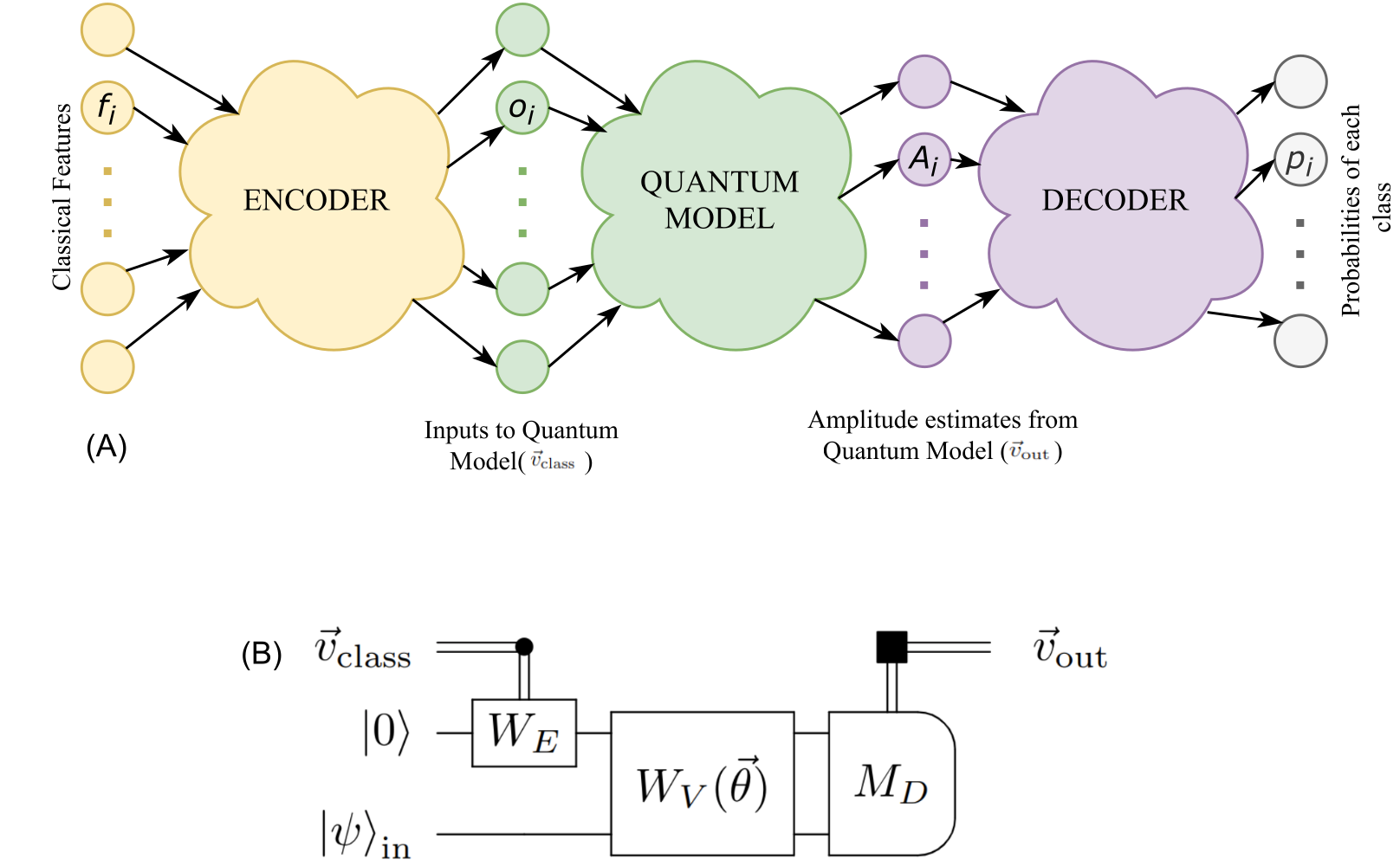}
	\caption{
		SQUID Model Architecture. A shows a current implementation, which is a simplified version of B, ($\ket{\psi}_{in} = 0$).
		In A the nodes can be thought of as neurons in classical machine learning, or quantum state vectors.
		The edges represent transformations applied to these states.
		These typically involve trainable weights, but they might, as often in case of Quantum Model perform a set of transformations defined by incoming data.
		"Inputs to the Quantum Model" ($\vec{v}_{class}$) in particular can define both the input state to a quantum system ($W_E$) along with the rotations that should be applied to that input state ($W_V(\theta)$).
		Quantum Model should also perform a measurement of the state ($M_D$) and return classical amplitude estimates ($\vec{v}_{out}$).
		Part B contains also planned future extension in which quantum features are also allowed to be passed in.
	}
	\label{fig:model-arch}
\end{figure}

The complete framework is shown graphically on Fig.~\ref{fig:model-arch}, and the detailed relations between models within the ensemble are described in the following subsections.

\subsection{Propagating through the Main Model}
\label{subsec:squid-main-model}

Calling the model, or calling the \texttt{forward} function is exactly the same to PyTorch's forward pass.
The only difference, is when middle model is Quantum, and the conversion between Tensors and numpy arrays~\cite{numpy} is required.
The reason for choosing numpy arrays to be passed into Quantum model is due to the fact that many QML packages accept them as the input and in fact prefer them even over standard Pythonic lists.

The \texttt{backward} function offers the only major modification for the user in comparison to PyTorch, and it is required to be called explicitly by the user.
For classical models standard PyTorch back-propagation (\texttt{autograd}) is used, and exact gradients are calculated.
In the case there is a Quantum model in the middle, the automatic gradient propagation stops at the end of the second (quantum) model.
This is because there was a conversion to/from numpy in the forward pass.
SQUID uses the \texttt{backward} function provided by implementation of the Quantum Model.
This both updates any parameters stored within the model, and provides the gradient with respect to the output of the encoder.

The conversion from numpy array to PyTorch tensor in backward call requires us to create a \textit{fake} loss, which we then use to propagate the gradients backward using back-propagation.
Given encoder forward output $o_{12}$, gradient of global loss with respect to that output $g_{12}$ (provided by the Quantum Model as described above),
we define a \textit{fake loss} $L'$:
\begin{align}
    L'(o_{12}, g_{12}) &= \sum_i\sum_j {o_{12}}_{ij}{g_{12}}_{ij} \\
    \frac{\partial L'(o_{12}, g_{12})}{\partial {o_{12}}_{ij}} &= {g_{12}}_{ij}
\end{align}
After the above calculation a PyTorch \texttt{backward} call is made on $L'$, which propagates the gradient using \texttt{autograd}.
Hence, gradients with respect to all of the Main Models parameters are calculated.
For clarity the process is shown in Algorithm~\ref{alg:main-backprop}.

\begin{footnotesize}
\begin{algorithm}[h]
	\SetAlgoLined
	\KwIn{Main Model $\mathcal{M}$, Loss $L$ (Torch Tensor)}

	Let $\mathcal{M}_1$, $\mathcal{M}_2$, $\mathcal{M}_3$ refer respectively to Encoder, Quantum Model, Decoder of $\mathcal{M}$\;

	L.backward()\;
	\If{$\mathcal{M}_2$ is Quantum}{
		Let be $g_{23}$ the gradient of $L$ with respect to input to $\mathcal{M}_3$ (PyTorch-provided)\;
		Let $g_{12}$ be the gradient of $L$ with respect to input to $\mathcal{M}_2$, this is achieved by passing $g_{23}$ through user-provided \texttt{backward} function of the $\mathcal{M}_2$ model\;
		Let $o_{12}$ be the output of $\mathcal{M}_1$ and input of $\mathcal{M}_2$ (Saved from forward iteration)\;
		Let $L'$ be sum of all elements of $g_{12} \odot o_{12}$\;
		$L'$.backward()\;
	}

	\caption{Back-propagation through the Main Model}
	\label{alg:main-backprop}
\end{algorithm}
\end{footnotesize}

\begin{footnotesize}
\begin{algorithm}[h]
	\SetAlgoLined
	\KwIn{Number of qubits $N$, vector of $M$ quantum parameters $\vec{\theta}$, desired number of outputs $Q_1\in[1,2^{N}]$}
	Construct unitary operation corresponding to quantum circuit\;
	Measure $Q_1$ output probabilities $p_k$ from circuit\;
	Use parameter shift rule and a total of $N_C=\min\left[2^{N}M,2Q_1(M+1)\right]$ circuits to estimate the $Q_1$ $M$-dimensional gradients $g_{kw}$\;
	\KwOut{Returns {\it result}=$\vec{p}$ and {\it grad}=$g_{kw}$}
	\caption{Quantum Model}
	\label{alg:quantum-backprop}
\end{algorithm}
\end{footnotesize}

It is worth mentioning that by using PyTorch built-in \texttt{autograd} procedure any PyTorch loss can be used.
For example, in the Section~\ref{sec:results} Cross Entropy loss is used.
Similarly any optimizer can be used.
The main caveat to using various optimizers is that if any parameters are defined within the Quantum Model the user has absolute control over updating them,
and the overhead of optimizer implementation falls onto the user.

\subsection{Complexity of Gradient Evaluation}
\label{sec:complexity}

A crucial question that needs to be evaluated to assess the practicality of any QML algorithm is the quantum complexity of the gradient calculation.  This is especially relevant since the gradients propagated through the quantum model require statistical sampling or a quantum technique such as amplitude estimation to evaluate~\cite{Brassard_2002,Suzuki_2020,Grinko_2021}.  Here we provide such a complexity analysis with the aim of bounding the scaling of the number of quantum operations needed to ensure that the gradients yielded by Algorithm~\ref{alg:quantum-backprop} are accurate within bounded error $\epsilon$.  Here in ensuring that the error is $\epsilon$ we mean that the gradient computation detailed in Algorithm~\ref{alg:quantum-backprop} outputs an estimate of the gradient $\widetilde{g}$ such that $|g-\widetilde{g}|_2 \le \epsilon$.

Let us consider a Monte-Carlo estimate of the gradient.  The algorithm for generating such an estimate involves measuring the expectation value of the gradient.  This expectation value can be evaluated using Hadamard tests to estimate each component of the gradient (see Appendix~\ref{app:gradients}).  Using the empirical frequency of measurements as an unbiased estimate of the probability, we have that if $\widetilde{g}$ is the estimate that returns from our protocol
\begin{equation}
| g - \mathbb{E}(\widetilde{g})|_2 =0. 
\end{equation}
As there are $M$ different parameters and $Q_1$ outputs yielded by the quantum model, we further have that
\begin{equation}
\mathbb{E}(|g - \widetilde{g}|_2^2) = \sum_{k=1}^{Q_1 M} \mathbb{E}(g_k^2 - 2 g\widetilde{g_k} +\widetilde{g_k}^2) =\sum_{k=1}^{Q_1 M} \mathbb{V}(\widetilde{g}_k).
\end{equation}
Since the variance of the sum is the sum of the variances and we are using the sample mean for our estimates of the probability.  This means that if $N$ samples are used per point then
\begin{equation}
\sum_{k=1}^{Q_1 M} \mathbb{V}(\widetilde{g}_k) \le \sum_{k=1}^{Q_1 M} \frac{\mathbb{V}(g_k)}{N} 
\end{equation} 
Therefore we have that the mean square error of $\widetilde{g}$ is at most $\epsilon^2$ if
\begin{equation}
N \geq \frac{Q_1M}{\epsilon^2} \max_{k}\mathbb{V}[g_{k}]\in O\left(\frac{Q_1M}{\epsilon^2}\right)\;,\label{eq:nsamp}
\end{equation}
note that if the variances are small then the number of samples required will be further reduced.
Then from Markov's inequality, the number of samples needed to estimate the gradient within error $\epsilon$ with probability greater than $2/3$ will be simply given by $3$ times the estimate in Eq.~\eqref{eq:nsamp} above. 

If a learning rate of $\lambda$ is used for the gradient ascent then the error in the quantum parameters (as quantified by the Euclidean distance) is, with probability greater than $2/3$, at most $\epsilon\lambda$.  We denote by $\Delta$ the error introduced by using the noisy estimator $\widetilde{g}$ for the parameter update
\begin{equation}
\Delta = \left\|\prod_j e^{-i \theta_{j=1}^M H_j} - \prod_j e^{-i \widetilde{\theta}_{j=1}^M H_j}\right\|_2\;,
\end{equation}
with $\|\cdot \|_2$ the induced Euclidean norm.
The generators $H_j$ used in SQUID (see Sec.~\ref{sec:qmodels}) have unit norm, this allows us to bound the error as
\begin{equation}
\Delta \le | \vec{\theta} - \vec{\widetilde{\theta}}|_1 \le \sqrt{M} | \vec{\theta} - \vec{\widetilde{\theta}}|_2 \in O\left({ \sqrt{M} \lambda \epsilon}\right)\;.
\end{equation}
Therefore it follows from the fact that the error in the output probabilities $\vec{p}$ satisfies $|p_k(\vec{\theta}) - p_k(\widetilde{\vec{\theta}})| \in O(\Delta)$, that the value of $N$ needed to ensure that the maximum error $\delta$ in $p_k$ after a single step of gradient ascent, with probability at least $2/3$, obeys
\begin{equation}
N \in O\left(\frac{Q_1 M^2 \lambda^2}{\delta^2} \right).
\end{equation}

Finally, an application of the Chernoff bound shows that if we wish the error to be $\delta$ with probability at least $1-\eta$ then we can repeat the experiment a logarithmic number of times and use majority voting to estimate the updated parameters.  This results in
\begin{equation}
N \in O\left(\frac{Q_1 M^2 \lambda^2}{\delta^2}\log\left(\frac{1}{\eta} \right) \right) \in \widetilde{O}\left(\frac{Q_1 M^2 \lambda^2}{\delta^2} \right),
\end{equation}
where $\widetilde{O}(\cdot)$ denotes an asymptotic upper bound with polylogarithmic multiplicative factors suppressed.
On a future fault-tolerant quantum device it would also be possible to obtain a quadratic speedup in both $\epsilon$ and $M$ at the cost of a longer circuit depth (see eg.~\cite{Brassard_2002,gilyen2019optimizing}).

\subsection{Available classical models}
There are two built-in classical models: Linear and Convolutional. Both accept arguments which specify numbers of neurons per layer, and activation functions between them.

However, since \texttt{ClassicalModel} is a sub-class of \texttt{nn.Module} from PyTorch, it is straightforward for a user to implement their own model.
This is recommended, unless configuration files from SQUID helpers are utilized (see Sec.~\ref{sec:squid_hlp}).

\subsection{Available quantum models}
\label{sec:qmodels}

In this section we provide details on the implementation of quantum models within SQUID. The approach we follow in this preliminary study is to construct the most general models on a given set of qubits by expressing the quantum circuits as layers of structured operations acting in nearest-neighbors only. This allows for both generality and a direct connection with real-world implementations on near-term devices with limited connectivity. Despite this choice, the framework is general and can be easily extended to accommodate quantum models with a different structure.

The common construction for a variational quantum classifier (see eg.~\cite{Havl_ek_2019,Schuld_2019b,Schuld2020}) is to start by considering the encoding of an input $D$-dimensional feature vector $\vec{v}\in[0,1]^{D}$  into the quantum state of a register containing $N\geq\lceil\log_2(D)\rceil$ qubit by introducing an encoding unitary operation $W_E$ as
\begin{equation}
\ket{\Psi\left(\vec{v}\right)} = W_E(\vec{v})\ket{0}\;,
\end{equation}
with $\ket{0}$ a reference state in the computational basis. The encoded state $\ket{\Psi}$ is then modified by acting with a second unitary $W_V$ defined in terms of a set of $N_v$ variational parameters $\vec{\theta}\in[0,\pi)^{N_v}$. The final state of the quantum register right before measurement is then
\begin{equation}
\label{eq:qfeat}
 \ket{\Phi\left(\vec{v},\vec{\theta}\right)} = W_V(\vec{\theta}) \ket{\Psi\left(\vec{v}\right)} = W_V(\vec{\theta})W_E(\vec{v})\ket{0}\;.
\end{equation}

The output of the quantum models we consider here are the probabilities of measuring each one of the computational basis states in the state $\ket{\Phi}$, which can be estimated by collecting statistics over a large number of circuit executions. Given that the number of possible outcomes scales exponentially in the register size, a small subset of probabilities is typically selected in order for the overall scheme to be scalable.

The decomposition of the total unitary operation mapping $\ket{0}$ to $\ket{\Phi}$ as a function an the {\it encoding} unitary $W_E$ and a {\it variational} unitary  $W_V$ is however artificial and does not necessarily lead to the most efficient scheme. This is especially true in the SQUID framework where a classical network is devoted to optimally determine an encoding of the classical data into a quantum state. The approach we take in SQUID is to consider instead the $M$-dimensional output from the classical encoder as the parameters describing a global unitary $W$ in the quantum register, without artificially distinguishing between ``encoding'' parameters and ``variational'' parameters. This hybridization of the standard approach described above is still completely general and the global network can adjust to effectively reproduce a factorized form $W=W_VW_E$ if there is a measurable advantage for the data under analysis.

We note that a generic unitary operation on $N$ qubits can depend on at most $(4^N-1)$ parameters, which implies that we need to choose $M<(4^N-1)$ for the output of the classical encoder. In practice this is not a limitation since unitary operations which can be decomposed efficiently into a polynomial number of one and two qubit operations will depend at most on a polynomially large number of parameters.

In this first exploratory study we use a simple, but general, parametrization of $W$ in terms of a one qubit unitary $U^1_{k}(\alpha,\beta,\gamma)$ and a two qubit unitary $U^2_{jk}(\theta,\phi,\eta)$ both parametrized by 3 real parameters taking values in $[0,\pi)$.
The unitary $U_1$ can be written as
\begin{equation}
\label{eq:u1_unitary}
U^1_k(\alpha,\beta,\gamma) = \exp\left(i\alpha Y_k\right)\exp\left(i\beta Z_k\right)\exp\left(i\gamma Y_k\right)\;,
\end{equation}
and we recognize the parameters $(\alpha,\beta,\gamma)$ to be the Euler angles in the $YZY$ decomposition.
In the expression above $Z_k$($Y_k$) denote the Pauli Z matrix (Y matrix) for qubit $k$, and we will denote $X_k$ similarly in the following.
The two-qubit unitary $U^2_{jk}$ acting on the qubit pair $\{j,k\}$ is instead defined as
\begin{equation}
\label{eq:u2_unitary}
U^2_{jk}(\theta,\phi,\eta) =e^{i\theta X_j\otimes X_k + i\phi Y_j\otimes Y_k + i\eta Z_j\otimes Z_k}\;.
\end{equation}

The usefulness of these choices comes from the possibility to represent a generic unitary by applying appropriately layers of $U^1_k$ and $U^2_k$ operations.
For instance, a general $SU(4)$ transformation for 2 qubits can be exactly represented with the following circuit (see~\cite{vidal2004,vatan2004})
\begin{equation*}
\label{circ:two_qubits}
\Qcircuit @C=1em @R=.7em {
& \gate{U^1_0(\alpha_0,\beta_0,\gamma_0)} & \multigate{1}{U^2_{01}(\theta,\phi,\eta)}&\gate{U^1_0(\alpha_2,\beta_2,\gamma_2)} & \qw\\
& \gate{U^1_1(\alpha_1,\beta_1,\gamma_1)} & \ghost{U^2_{01}(\theta,\phi,\eta)}&\gate{U^1_1(\alpha_3,\beta_3,\gamma_3)} & \qw\\
}
\end{equation*}
requiring $1$ application of $U^2_{01}$ and $4$ applications of $U^1_k$ (on qubit 0 and 1) for a total of $15$ parameters.
This construction can be readily extended to larger systems by interleaving layers of $U^1_k$ with layers of $U^2_{jk}$ applied alternatively on even or odd partitions.
For instance with $4$ qubits we consider circuits of the following form
\begin{equation*}
\label{circ:four_qubits}
\Qcircuit @C=1em @R=.7em {
& \gate{U^1_0} & \multigate{1}{U^2_{01}}&\gate{U^1_0} & \qw                                       &\qw& \multigate{1}{U^2_{01}}&\gate{U^1_0}& \qw\\
& \gate{U^1_1} & \ghost{U^2_{01}}              &\gate{U^1_1} & \multigate{1}{U^2_{12}}&\gate{U^1_1} & \ghost{U^2_{01}}&\gate{U^1_1}& \qw\\
& \gate{U^1_2} & \multigate{1}{U^2_{23}}&\gate{U^1_2} & \ghost{U^2_{12}}             &\gate{U^1_2}&\multigate{1}{U^2_{23}}&\gate{U^1_2}& \qw\\
& \gate{U^1_3} & \ghost{U^2_{23}}             &\gate{U^1_3} & \qw                                        &\qw&\ghost{U^2_{23}}&\gate{U^1_3}& \qw\\
}\;.
\end{equation*}

Note that in the construction above we didn't include the first and last single-qubit operations in the third $U^1$ layer. This allows to remove redundancy in the parameters since we replace the product of two $U^1$ operations with a single $U^1$, this simplification results in enhanced stability in the training.

\section{Example applications}
\label{sec:results}
As an initial application of our framework, we present here results for binary classification on the MNIST database~\cite{MNIST} using digits 3 and 7. This is a standard benchmark for classification algorithms and analysis with a quantum model is made possible by the ability of SQUID to compress the input features into data with the appropriate dimensions.

\begin{figure}[]
	\centering
	 \includegraphics[width=0.45\textwidth]{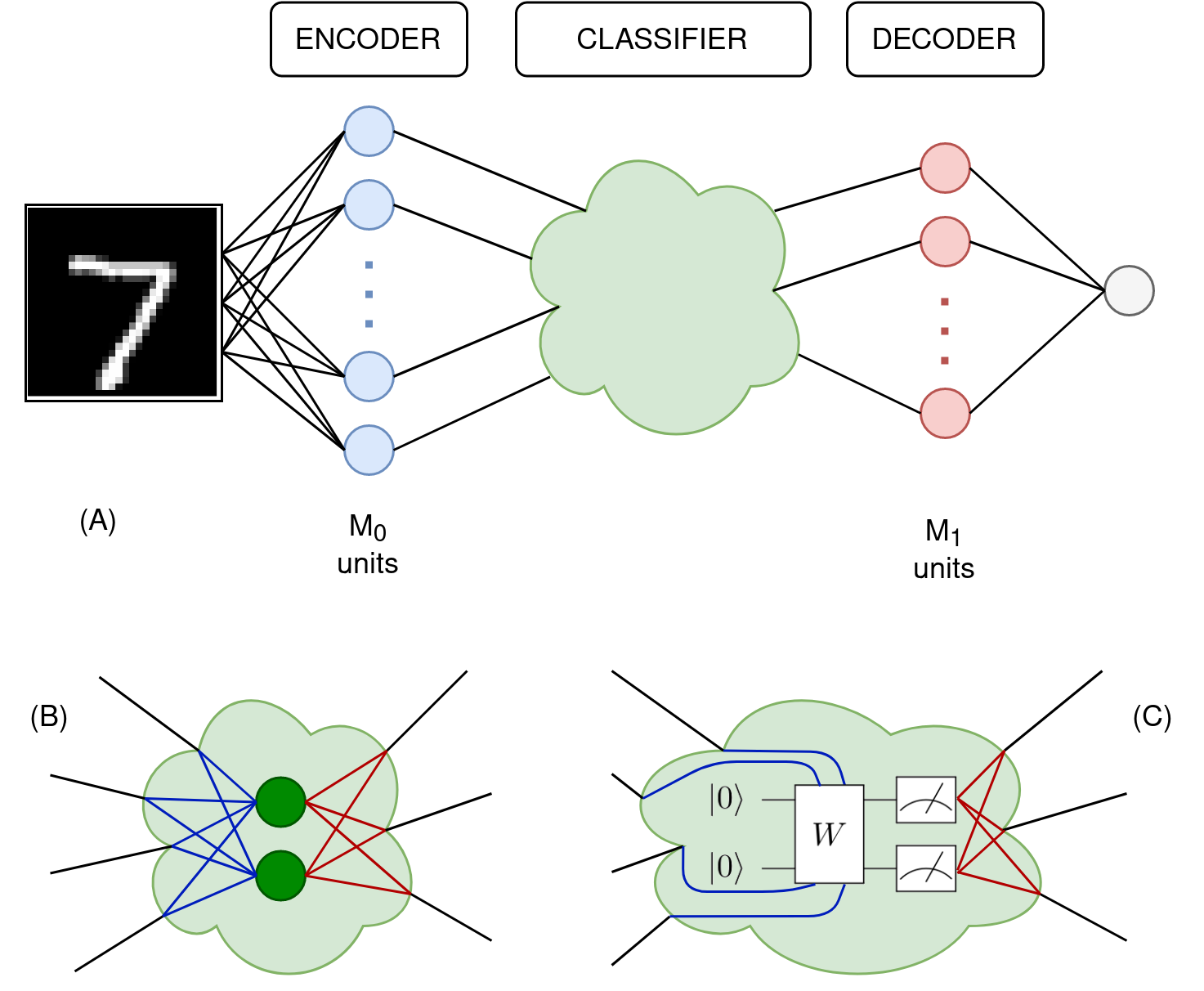}
	\caption{ Pictorial representation of the hybrid classifier model used for MNIST classification. Panel $(A)$ shows the complete network including an encoder with $M_0$ units and a decoder with $M_1$ units. Panel $(B)$ shows the implemented classical classifier composed by two units and panel $(C)$ shows a schematic of the quantum models: input parameters coming from the encoder determine the unitary $W$ while the output is obtained upon measurement of the qubits. }
	\label{fig:mnist_model}
\end{figure}

The input feature vectors for this dataset are real vectors of dimension 784 representing a grayscale $28\times 28$ pixel picture.
In this section we will compare results obtained with the architecture displayed in Fig.~\ref{fig:mnist_model} composed by: a single layer encoder with $M_0$ units, a black-box classifier to be specified below and a single layer decoder with $M_1$ units.
The classical black-box classifier used in this section consists of a simple single layer network with $2$ units (panel $(B)$ of Fig.~\ref{fig:mnist_model}) and we take $M_1=2$ for it's output.
In the following subsections we will also consider different implementations of quantum classifiers with the general structure displayed in panel $(C)$ of the same figure.

All of the calculations (classical and quantum) presented in this section were obtained using an Adam optimizer as implemented in PyTorch and using the hyperparameters reported in Tab~\ref{tab:hyppar}. In all case we use $48$ independent optimization runs that were performed in order to estimate the variance in the attained accuracy. In the following we will refer to this ensemble as "Bootstrap runs".

\begin{table}[]
\footnotesize
\centering
\begin{tabular}{l|l|l|l|l}
Batch size & Epochs & Learning rate & Train size & Val. size\\ \hline
16 & 100 & 0.0001 & 9916 & 2480
\end{tabular}
\caption{Hyperparameters used for the results on MNIST. \label{tab:hyppar}}
\end{table}

\begin{figure}[]
	\centering
	 \includegraphics[width=0.45\textwidth]{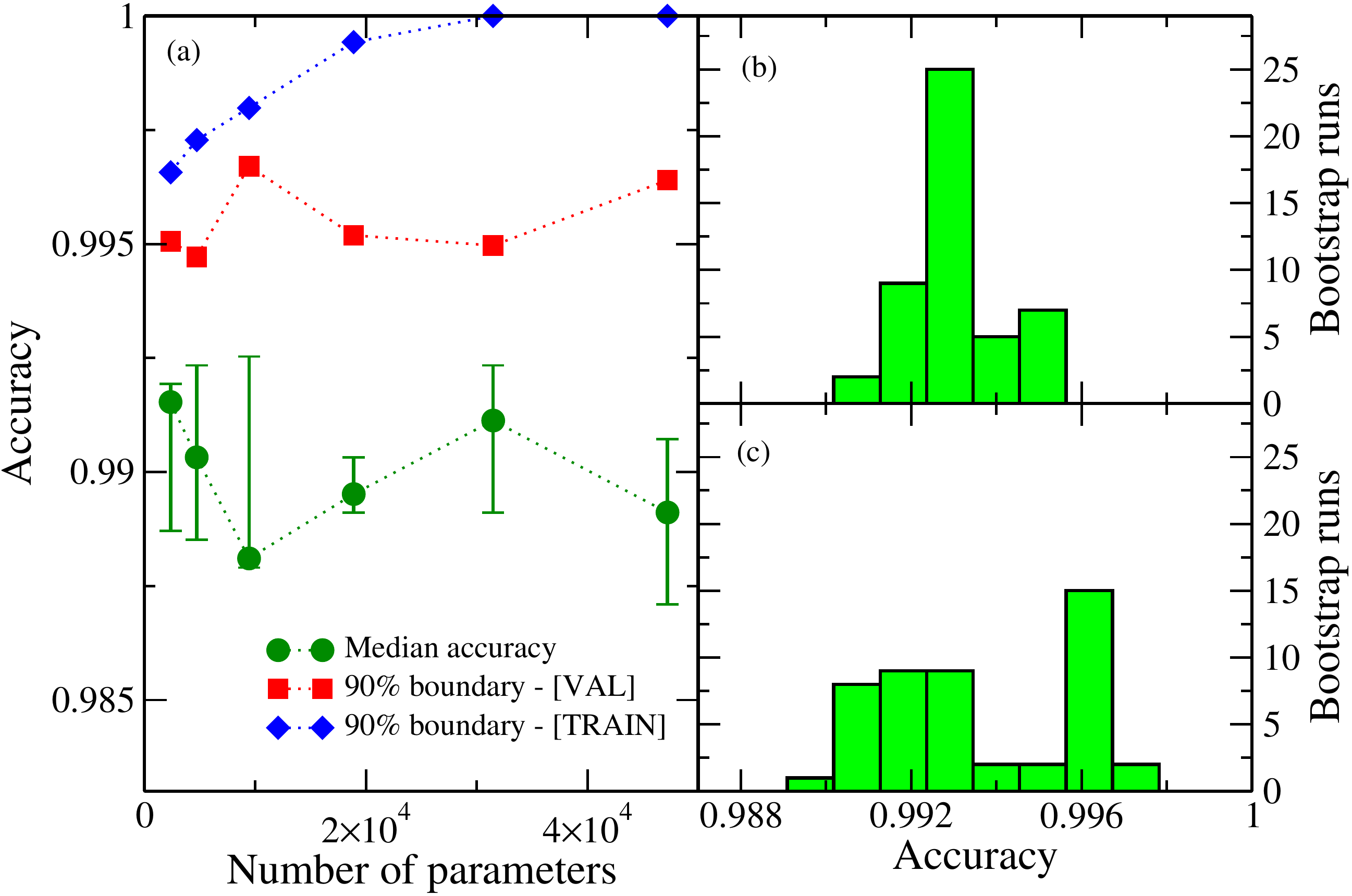}
	\caption{ Results for the accuracy achieved on MNIST with the classical model described in the text. Panel (a) shows the accuracy as a function of the number of parameters in the model (green solid circles), the other two sets of points show the location of  the $90\%$ accuracy percentile for both the validation set (red squares) and the training set (blue diamonds) . Panels (b) and (c) show the histogram of achieved accuracies for the smallest ($M_0=3$) and largest ($M_0=60$) models respectively.  }
	\label{fig:cl_accuracy}
\end{figure}

We present the results obtained with the classical network in Fig.~\ref{fig:cl_accuracy} (the full data is available in Tab.~\ref{tab:cl_res} of Appendix~\ref{app:mnist}). We can see from the evolution of the median accuracy (green circles) that the classical network is able to achieve classification accuracies above $99\%$ but the increase in the number of hidden units at the level of the input model doesn't seem to provide a statistically significant improvement on the final accuracy. The displayed error bars are $68\%$ confidence intervals extracted from our finite population sample.

In the main panel, we also show the location of the $90\%$ accuracy percentile, ie. the boundary value for which $10\%$ of bootstrap runs provide a higher accuracy, for both the validation set (red squares) and the training set(blue diamonds).
These results are consistent with the expectation that, as the number of parameters $K_{tot}$  in the model increases, the training set can be described almost exactly by the network while at the same time we see that the distribution of accuracies for the  validation set does not evolve significantly with $K_{tot}$. In order to understand this point better we show in the left panels the estimated histograms for accuracy reached in our set of $48$ bootstrap runs for the smallest classical model (panel (b)) and the largest model (panel (c)). As expected from the results in the main panel, most of the density is in the same location but for the larger models the tails are more important.

\begin{figure}[]
	\centering
	 \includegraphics[width=0.45\textwidth]{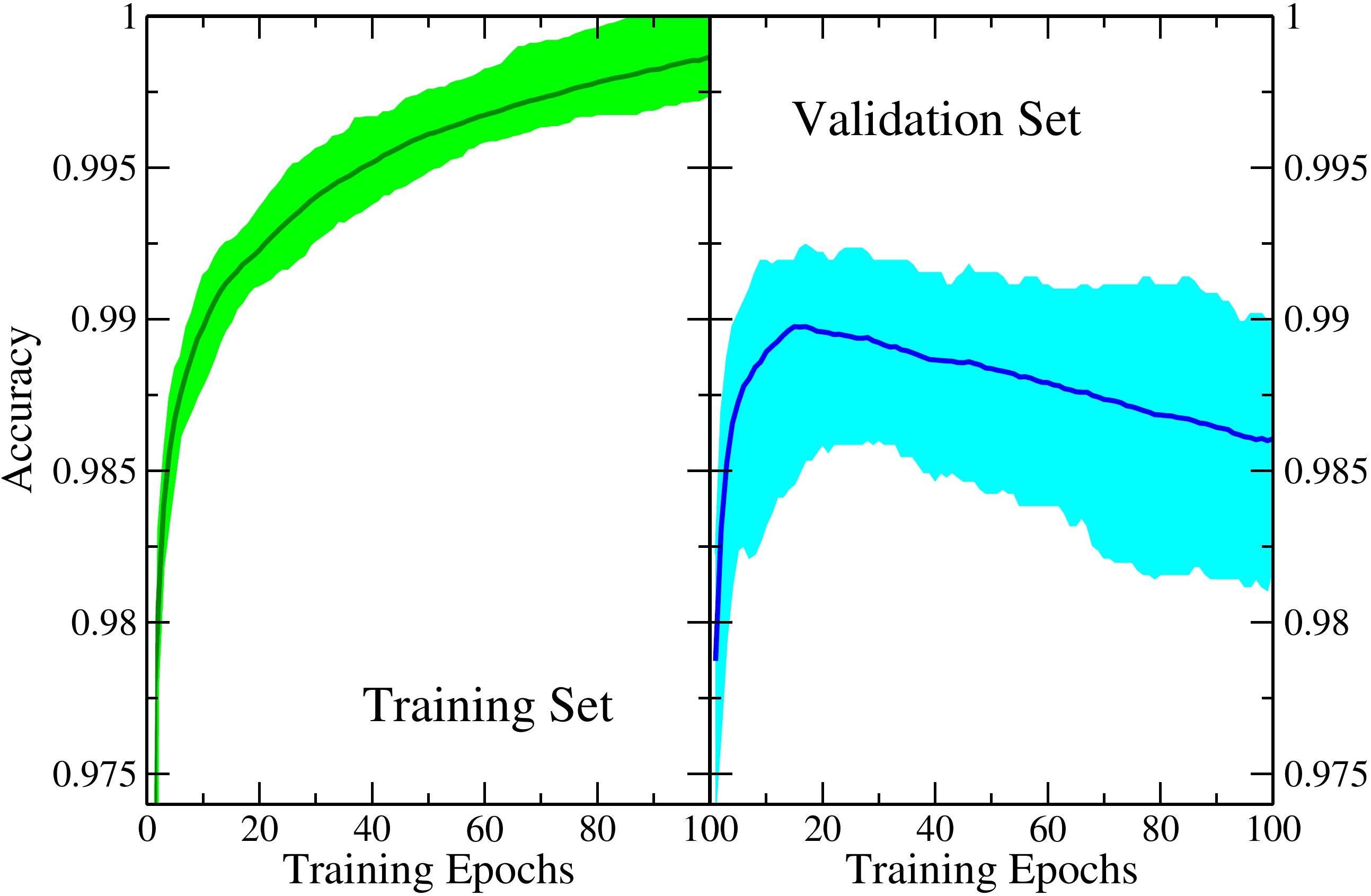}
	\caption{ Example of training of the classical model described in the main text. The left panel shows the increase in classification accuracy for the training set as a function of the number of epochs. The right panel shows the same for the validation set.  All bands are $90\%$ confidence intervals with averages indicated by the solid lines.}
	\label{fig:cl_train}
\end{figure}

Note that the dispersion in the accuracy around the median reported in the main panel of Fig.~\ref{fig:cl_accuracy} is relatively small, of the order of $\approx0.002$. This is caused in large part by the simplicity of the classification problem, as we can see in Fig.~\ref{fig:cl_train} (obtained for a medium sized model with $M_0=40$) the accuracy quickly converges to a narrow interval around the mean for both the training set and the validation set. With more than $~80$ epochs the accuracy for the training data set is able to reach $100\%$.

When the inner model is replaced by a quantum subroutine, as depicted in panel $(C)$ of Fig.~\ref{fig:mnist_model}, the output dimension for a quantum circuit over $N$ qubits is bounded by $M_1\leq2^{N}$. In the following sections we will consider two limiting situation: the maximum possible dimension $M_1=2^{N}$ (indicated as ``full'' below) and the minimum one $M_1=2$ (indicated as ``min'' below) and corresponding to the probability of measuring a single basis state (here we choose $\ket{0}^{\otimes N}$).

\subsection{Separable Quantum Models}

The first class of quantum models we consider here are separable models with a single layer of $U^1$ unitaries and are therefore fundamentally classical in that entanglement plays no role in shaping the output probabilities of the quantum model. The results are shown in Fig.~\ref{fig:qm_sep_accuracy} and the full data is available in Tab.~\ref{tab:qm_res_A} of Appendix~\ref{app:mnist}.

\begin{figure}[]
	\centering
	 \includegraphics[width=0.45\textwidth]{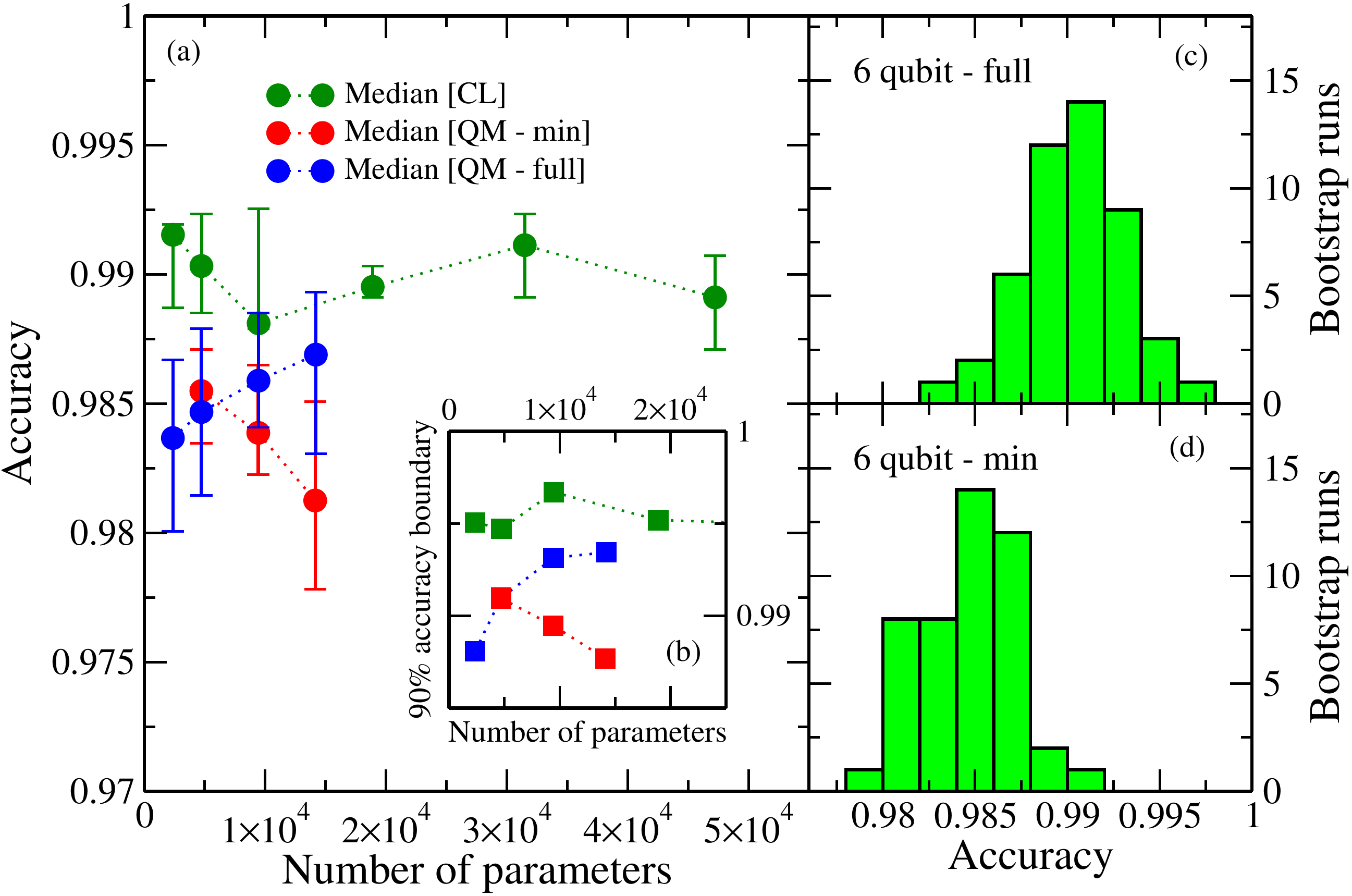}
	\caption{ Results for the accuracy achieved on MNIST with the classical model (green points) and the separable quantum models described in the text (red and blue points). Panel (a) shows the accuracy as a function of the number of parameters for the classical model (green solid circles), the full quantum separable models (blue solid circles) and the separable quantum models with a single output variable $M_1=1$ (red solid circles) indicated by ``min''. The inset panel (b) shows the location of the $90\%$ accuracy percentile for the classical (green squares), full separable quantum model (blue squares) and minimal separable quantum model (red squares). Panels (c) and (d) show the histogram of achieved accuracies for the 6 qubit models with either the full number of possible output variables ($M_1=64$, top panel) and the single output (bottom panel).  }
	\label{fig:qm_sep_accuracy}
\end{figure}

In this case, at least for the models with $M_1=2^{N}$, we can see a mild increase of the final accuracy as a function of the number of parameters/qubits in the model, the largest model is however outperformed by classical networks with a smaller size (see the blue circles in Fig.~\ref{fig:qm_sep_accuracy}(a)). The models with a latent space corresponding to the restricted output for the quantum layer (shown as red circles in Fig.~\ref{fig:qm_sep_accuracy}(a)) show instead a deterioration as we increase the number of qubits/parameters. This effect is especially clear looking at the histograms of achieved accuracy in the 48 bootstrap runs displayed in the right panels of Fig.~\ref{fig:qm_sep_accuracy} for the largest model with $N=6$ qubits: employing a large output vector from the quantum layer produces results with better than $99\%$ for more than half of the runs while restricting the output to a single probability prevents most runs from reaching this threshold. Strikingly, this is true even for the training set (not shown) where only a single bootstrap run achieved an accuracy above $99\%$. This is a first clear sign of the importance to supplement the quantum classifier with a rich decoder at the possible expense of a larger sample complexity.

\subsection{Quantum models with entanglement}
\label{sec:qm_res_withent}

\begin{figure}
	\centering
	 \includegraphics[width=0.45\textwidth]{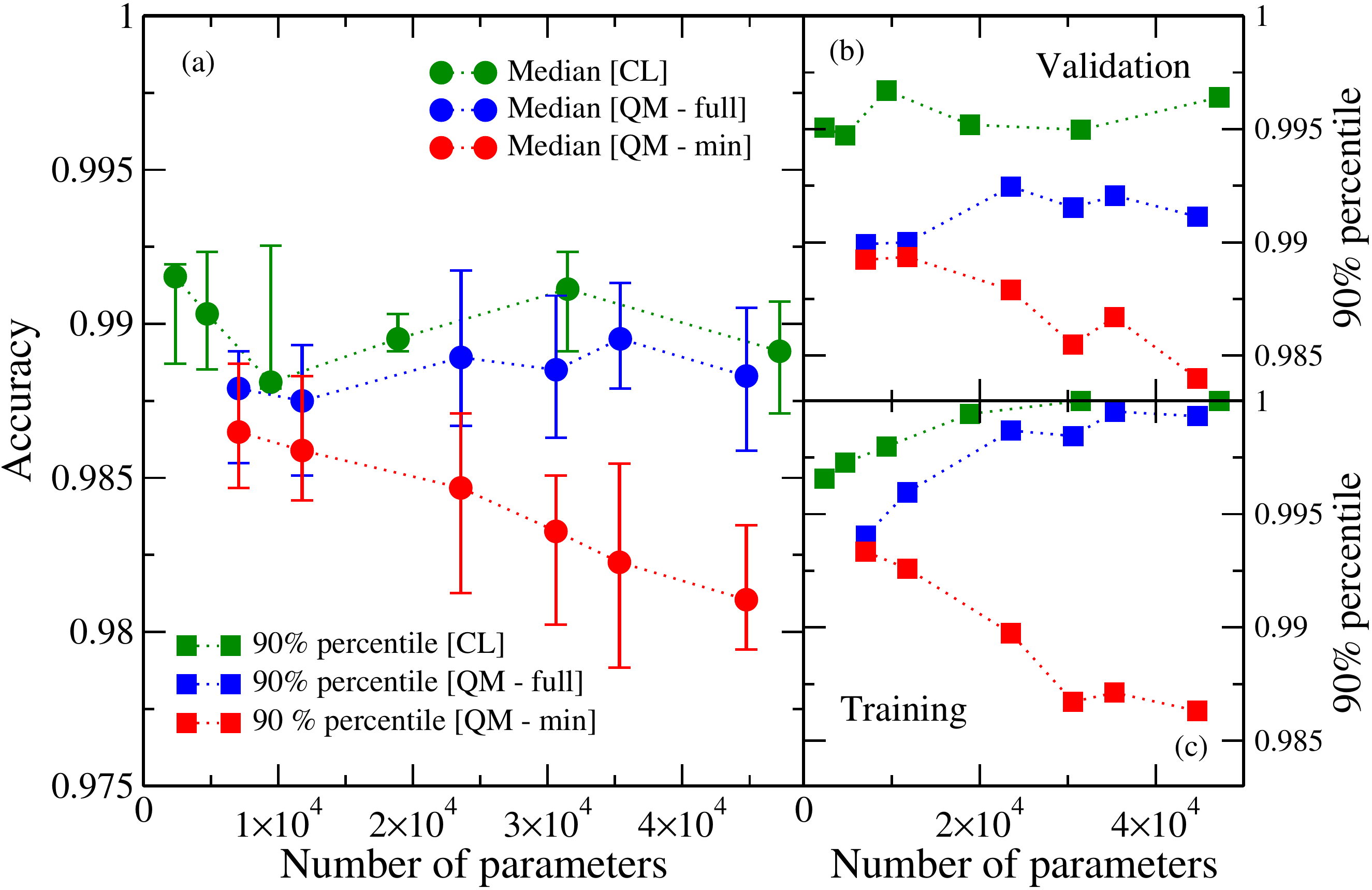}
	\caption{ Results for the accuracy achieved on MNIST with the classical model (green points) and the quantum models with entanglement described in the text (red and blue points). Panel (a) shows the accuracy as a function of the number of parameters for the classical model (green solid circles), the full quantum models (blue solid circles) and the quantum models with a single output variable $M_1=1$ (red solid circles) indicated by ``min''. The left panels show the location of the $90\%$ accuracy percentile for the classical (green squares), full quantum (blue squares) and minimal quantum model (red squares) in either the validation set (panel (b)) or the training set (panel (c)).	\label{fig:qm_accuracy_bounds}}
\end{figure}

We now turn to consider more general quantum models that are capable of creating entanglement in the quantum register through the use of the two-qubit unitary $U^2$ defined in Sec.~\ref{sec:qmodels}. The resulting median accuracy shown in panel (a) of Fig.~\ref{fig:qm_accuracy_bounds} show a similar trend to the simpler separable models above: the accuracy of the quantum model never exceeds the classical accuracy and there doesn't seem to be any measurable advantage in increasing the number of parameters. Interestingly, for the models with restricted output size (red circles denoted $QM - min$ in panel (a) of Fig.~\ref{fig:qm_accuracy_bounds}), we see that the optimization procedure is struggling to find a good set of parameters for the larger models and the accuracy decreases almost monotonically with size. It is possible that better results could be obtained using directly the accuracy as cost function instead of the cross-entropy. In order to clarify that the effect we are seeing is not coming from over-fitting of the training set, but really from difficulties in exploring efficiently the energy surface, we show on the left panels the evolution of the $90\%$ accuracy percentile as a function of the number of parameters for both the validation and training set (panels (b) and (c) respectively) for the 3 networks considered here: the classical feed-forward network considered before (green squares) and the quantum models with entanglement either with the full output (red squares) or the restricted output model (blue squares). As can be clearly seen in panel (c) the optimizer is not able to find a good parameter set for large models and the accuracy in training decreases.

These results highlight the importance of supplementing the quantum classifier with a non trivial output decoder in order to achieve a good efficiency, a possibility that is available only if we choose to measure more than a single qubit from the quantum device.

\section{SQUID Extensions and Future Work}
\label{sec:ext}
As mentioned is Section~\ref{sec:squid} SQUID is designed with extensibility in mind.
This means that it should be easy to write additional packages on top of it, as well that additional features should involve changing at most few modules. In the following two sections we describe the SQUID Helpers package, designed to allow the use of SQUID using user-provided configuration files, and comment on possible extensions of the framework.

\subsection{SQUID Helpers Package}
\label{sec:squid_hlp}
To show how such extension could function, but also to stream-line the workload for use cases,
we provide SQUID Helpers extension.
It allows user to use yaml configuration files to run SQUID code.
As a result bootstrapping multiple runs of multiple test cases and aggregation of the results is much easier.

The conversion from configuration files to SQUID is done when a file is first read, and if a bootstrap option is provided, then random seed is changed during every iteration.
The change of the seed is deterministic, and hence all of the results are exactly reproducible.
At the initialization step, there is a single batch forward pass of data to ensure that dimensions of models inputs and outputs match.
This is done due to fail-first and fail-fast principle.
A single forward pass of data is faster than a pass of a single batch, and in perspective of training it is very cheap.
The code then runs a typical training for-loop, with an additional call to \texttt{backward} function of Main Model, as explained in Section~\ref{subsec:squid-main-model}.
In the end, all of the results, as well as configuration is saved to a single location.
If bootstrap was used, along with the results for each run, there is a folder with aggregated results created for simpler analysis.

Helpers extension also provides very basic plotting utilities for the results.
However, those are meant as a example of processing the output folders, since plots are highly dependent on studies performed.

\subsection{Future Work}

There is a vast amount of possible extensions to SQUID, some of which can be included directly in a main project, while others can be used as standalone packages.
The main advantage of the SQUID is that it allows for abstracting communication with a specific backend.
To do so however, custom \texttt{QuantumModel} subclasses interfacing with the backend API will be needed.

Additionally, as of right now SQUID allows only for a classification and regression tasks.
More advanced scenarios would require changes both in SQUID as well as, to a larger extent, SQUID helpers.

Another, and much sooner addition to SQUID will be to start supporting various quantum computing frameworks.
For example, Qiskit has created a great package for optimization of quantum circuits~\cite{Qiskit}.
This would fit perfectly into SQUID ecosystem with a translation layer.
Such addition would allow the user to implement hybrid models without the explicit definition of circuits for training, provided the circuit gradient are correctly passed by the backward function.

There are also others frameworks that have either already similar behavior or plans for optimization packages.
Modularity of SQUID would allow code to be backend agnostic, and work uniformly across multiple types of quantum devices.

\section{Conclusions}
\label{sec:conclusions}

The great success of classical machine learning algorithms in tasks such as classification, together with the expectation that quantum computers will allow us to explore algorithms in a larger complexity class than their classical counterparts, makes the exploration of the connections between these ideas a fertile ground for the discovery of novel approaches to automated inference.

In this work we have presented SQUID as a computational framework which allows to explore efficiently the possible advantages of quantum computing for machine learning purposes. This is achieved by embedding the quantum algorithm part in a more general multi-layer architecture that allows to interface classical and quantum networks while enjoying efficient optimization by using the automatic differentiation engine provided by PyTorch.
While there are similar packages (notably Xanadu's PennyLane~\cite{Bergholm2020Pennylane}, IBM's Qiskit~\cite{Qiskit}), they do not offer as much flexibility as SQUID.
For example, PennyLane offers much more complete experience, with many examples and models, as well as ability to run code on quantum computer, and not a simulator.
However, for the same reason, creating a custom model is much easier in SQUID.
Similar argument can be made about Qiskit.
This generalized frameworks provides several advantages over an either purely classical or purely quantum approach: it allows for a seamless dimensionality reduction of the inference problem, a step that would be necessary to explore high dimensional datasets on small quantum devices, while at the same time allowing for automatic tuning of the measurement settings needed to extract information from the quantum state produced by the algorithm.
This latter feature, implemented in SQUID by using a classical network as decoder after the central model as shown in Fig.~\ref{fig:model-arch}, is extremely important in order to reach high precisions.
The use of an explicit decoder at the output of the quantum model allows for a more careful optimization of the trade-off between measuring only vanishingly small fraction of the possible output probabilities on one hand, with the drawback that entanglement can start to be detrimental due to information scrambling (see eg.~\cite{Shen2020,marrero2021entanglement}), and a full measurement of the probability distribution on the computational basis states which will require an exponential number of repetitions.
In this work we used a simple classification problem from the MNIST database to show the effect of this tradeoff for a concrete high-dimensional problem.

Thanks to the generality of the architecture developed in this work, future explorations of algorithms with entanglement but with a classically efficient representation (such as tensor network states with polynomially large bond dimension, see eg.~\cite{Liu_2019,roberts2019tensornetwork}) could be carried out within SQUID with only minimal modifications to the code.
We expect the added flexibility in interchanging classical and quantum components in a global classifier to prove valuable in identifying promising datasets and inference problems where the presence of entanglement and quantum correlations can provide important accuracy gains.
Once these problems have been identified with a simplified approach as the one currently implemented in SQUID, a successive study of the sample complexity along the lines of the derivation presented in Sec.~\ref{sec:complexity} will be needed to assess the practical viability of the algorithm.
Extensions to implement the effect of finite statistics, together with more advanced effects such as models of decoherence for a specific target device, can be added efficiently within SQUID and we plan to explore their impact on classification problems in future work.

\begin{acknowledgements}
The work of A. Roggero was supported by the InQubator for Quantum Simulation under U.S. DOE grant No. DE-SC0020970 and by the Institute for Nuclear Theory under U.S. Department of Energy grant No. DE-FG02-00ER41132. 
The work of J. Filipek was supported in part by a Washington Research Foundation Fellowship at the University of Washington.
The work of S.-C. Hsu is supported by the U.S. Department of Energy, Office of Science, Office of Early Career Research Program under Award number DE-SC0015971. Support for Nathan Wiebe was provided by the Laboratory Directed Research and Development Program at Pacific Northwest National Laboratory, a multi-program national laboratory operated by Battelle for the U.S. Department of Energy, Release No. PNNL-SA-157287 and the theoretical work on this project by NW was supported by the U.S. Department of Energy, Office of Science, National Quantum Information Science Research Centers, Co-Design Center for Quantum Advantage under contract number DE-SC0012704. Additional support for Nathan Wiebe was provided by Google Research Award.
\end{acknowledgements}

\appendix
\section{Gradient evaluation for variational quantum models}
\label{app:gradients}

In this appendix we provide a derivation of the scheme employed in the main text to evaluate gradients with respect to the parameters of the quantum models. Following the discussion in Sec.~\ref{sec:qmodels} of the main text, a generic variational quantum model is defined by a unitary transformation $W(\vec{w})$ dependent on $M$ real parameters $\vec{w}$ taking values in $[0,\pi)$. The general construction employed in this work consists in decomposing the full unitary operation $W(\vec{w})$ into a combination of one and two qubit unitaries denoted by $U^1_k(\alpha,\beta,\gamma)$ and $U^2_{jk}(\theta,\phi,\eta)$ respectively. The subscript indices indicate the qubit (or pairs of qubits) the unitary acts upon. Given the simple structure of a generic circuit as the one depicted in Eq.~\eqref{circ:four_qubits}, one can obtain a closed form expression for every component of the gradient by looking at the individual gradients of the basic unitaries $U^1$ and $U^2$ directly. Here we will use as a concrete example the generic $SU(4)$ unitary transformation from Eq.~\eqref{circ:two_qubits} which we reproduce here for convenience
\begin{equation*}
\Qcircuit @C=0.5em @R=.7em {
& \gate{U^1_0(\alpha_0,\beta_0,\gamma_0)} & \multigate{1}{U^2_{01}(\theta,\phi,\eta)}&\gate{U^1_0(\alpha_2,\beta_2,\gamma_2)} & \qw\\
& \gate{U^1_1(\alpha_1,\beta_1,\gamma_1)} & \ghost{U^2_{01}(\theta,\phi,\eta)}&\gate{U^1_1(\alpha_3,\beta_3,\gamma_3)} & \qw\\
}\;.
\end{equation*}
Note that in this case $M=15$. 
In the following we will assume, without loss of generality, that this unitary operation is applied to the initial state $\ket{00}$ of the two qubit register and denote the resulting state by
\begin{equation}
\rvert \Phi\left(\vec{w}\right)\rangle = W\left(\vec{w}\right)\ket{0}\;.
\end{equation}
The classical output generated by a measurement on the qubit register can be completely characterized by the $4$ probabilities to find the system in each one of the possible basis states:
\begin{equation}
\label{eq:out_prob}
p_k = Tr\left[\Pi_k\rvert  \Phi\left(\vec{w}\right)\rangle\langle\Phi\left(\vec{w}\right)\lvert \right]
\end{equation}
where we have introduced explicitly the projectors
\begin{equation}
\begin{split}
\Pi_0 &= \rvert00\rangle\langle00\lvert\quad\Pi_1 = \rvert01\rangle\langle01\lvert\\
\Pi_2 &= \rvert10\rangle\langle10\lvert\quad\Pi_3 = \rvert11\rangle\langle11\lvert\;.
\end{split}
\end{equation}

Computing the derivatives with respect to the 12 angles corresponding to the 4 one qubit $SU(2)$ operations is straightforward by recalling the definition Eq.~\eqref{eq:u1_unitary} of $U^1$ in terms of exponentials of Pauli operators. As an example, the derivative with respect to $\gamma_0$ of any of the probabilities in Eq.~\eqref{eq:out_prob} can be written explicitly as
\begin{equation}
\begin{split}
\label{eq:gradient}
\frac{\partial}{\partial \gamma_0} p_k =& \frac{\partial}{\partial \gamma_0} \langle00\lvert W^\dagger(\vec{w})\Pi_kW(\vec{w})\rvert00\rangle\\
=&i\langle00\lvert W^\dagger(\vec{w})\Pi_kW(\vec{w})Y_0\rvert00\rangle\\
&-i\langle00\lvert Y_0W^\dagger(\vec{w})\Pi_kW(\vec{w})\rvert00\rangle\\
=&-2\mathcal{I}\left[\langle00\lvert W^\dagger(\vec{w})\Pi_kW(\vec{w})Y_0\rvert00\rangle\right]\\
=&2\mathcal{R}\left[\langle00\lvert W^\dagger(\vec{w})\Pi_k\frac{\partial W(\vec{w})}{\partial \gamma_0}\rvert00\rangle\right]\;,
\end{split}
\end{equation}
where $\mathcal{I}$ ($\mathcal{R}$) indicating the imaginary (real) part. Note that, for all of the 12 parameters characterizing the single qubit transformations, the derivative of the full variational circuit unitary $W$ can be expressed in terms of the same parametrized unitary with the appropriate angle angle shifted by $\pi/2$. For the case of $\gamma_0$ considered above we have for instance:
\begin{equation}
\frac{\partial W(\vec{w})}{\partial \gamma_0} = i W(\vec{w})Y_0 = W\left(\vec{w}'(\gamma_0)\right)
\end{equation}
with a new set of parameters given by
\begin{equation}
\begin{split}
\vec{w}'(\gamma_0) = \bigg(&\alpha_0,\beta_0,\gamma_0+\frac{\pi}{2},\alpha_1,\beta_1,\gamma_1,\\
&\theta,\phi,\eta,\\
&\alpha_2,\beta_2,\gamma_2,\alpha_3,\beta_3,\gamma_3\bigg)\;.
\end{split}
\end{equation}
Using the optimal implementation for the more general $SU(4)$ transformation derived in Ref.~\cite{vatan2004} (see Fig.6 there) one can show that we have the same property for the 2 qubit unitary $U^2_{jk}$. This property is usually referred to as the parameter shift rule~\cite{Mitarai_2018,Schuld_2019}.

In order to estimate the expectation values in the last line of Eq.~\eqref{eq:gradient} we can employ two strategies: if the required number of output probabilities $K$ is the maximum possible one with $n$ qubits (ie. $K=2^n$), it is convenient to first decompose the projectors in the computational basis states into a linear combination of $K=2^n$ diagonal operators obtained by considering all the possible tensor products of identities and Pauli Z, and then to evaluate each one of the resulting expectation values using a single Hadamard test each. The total number of separate circuits required for this approach is then $KM$, with $M$ the total number of parameters.

In the more realistic situation where $K\ll2^n$ instead, the strategy just described will still require an exponential number of measurement in the size of the qubit register. A more efficient alternative can be obtained by evaluating explicitly the $K$ pairs of expectation values
\begin{equation}
r_k = \mathcal{R} \left[\langle 00 \lvert W^\dagger(\vec{w}) \rvert k\rangle\right]\quad i_k = \mathcal{I} \left[\langle 00 \lvert W^\dagger(\vec{w}) \rvert k\rangle\right]\;,
\end{equation}
with $\rvert k\rangle$ the computational basis state associated to the projector $\Pi_k$. These expectation values can be estimated using an Hadamard test with one additional ancilla qubit and require the execution of $2K$ independent circuits (one each for real and imaginary part).

For each one of the $M$ parameters, we then use additional $2K$ Hadamard tests to estimate the expectation values associated with the shifted unitaries
\begin{equation}
\widetilde{r}_{km} = \mathcal{R} \left[\langle k \lvert W(\vec{w})_m' \rvert 00\rangle\right]\quad \widetilde{i}_{km} = \mathcal{I} \left[\langle k \lvert W(\vec{w})_m' \rvert 00\rangle\right]\;,
\end{equation}
where we used the compact notation $W(\vec{w})_m'$ to indicate the derivative with respect to the $m$-th parameter.
This requires a total of $2KM$ independent circuit executions for a total of $2K(M+1)$ observables. The gradient can then be computed as
\begin{equation}
\begin{split}
\label{eq:gradient_Htest}
\frac{\partial}{\partial w_m} p_k =&2\mathcal{R}\left[\langle00\lvert W^\dagger(\vec{w})\Pi_kW(\vec{w})_m'\rvert00\rangle\right]\\
=&2\left(r_k\widetilde{r}_{km}- i_k\widetilde{i}_{km}\right)\;.
\end{split}
\end{equation}

An alternative approach to reduce the number of independent circuits needed for gradient evaluation is to use expectation values of unitary operators instead of projectors. This extension can be easily implemented within the SQUID framework.

\section{Additional information on the MNIST benchmark}
\label{app:mnist}

\begin{table}[]
\centering
\begin{tabular}{l|l|l|l|l|l}
model & $M_0$ & $K_{tot}$ & Accuracy & TR$_{90}$ & VR$_{90}$\\ \hline
$cA$ & 3 & 2366 & $0.9915^{+4}_{-28}$ & 0.9966 & 0.9951\\
$cB$ & 6 & 4727 & $0.9903^{+20}_{-18}$ & 0.9973 & 0.9947 \\
$cC$ & 12 & 9449 & $0.9881^{+44}_{-2}$ & 0.9980 & 0.9967 \\
$cD$ & 24 & 18893 & $0.9895^{+8}_{-4}$ & 0.9994 & 0.9952 \\
$cE$ & 40 & 31485 & $0.9911^{+12}_{20}$ & 1 & 0.9950 \\
$cF$ & 60 & 47225 & $0.9891^{+16}_{20}$ & 1 & 0.9964
\end{tabular}
\caption{Results for the classical feed-forward models described in the main text. \label{tab:cl_res}}
\end{table}

We report in Tab.~\ref{tab:cl_res} the parameters and results for the classical models used in the MNIST classification discussed in Sec.~\ref{sec:results} and corresponding to the results presented in Fig.~\ref{fig:cl_accuracy} on the main text.
The last two columns in Tab.~\ref{tab:cl_res} denoted by TR$_{90}$ and VR$_{90}$ show the boundary value for the $90\%$ accuracy percentile, the latter refers to the validation data while the former to the training set. The estimated errors correspond to a $68\%$ confidence interval.

The parameters of the separable quantum models considered in the main text, together with the results obtained from training on the MNIST classification problem, are presented in Tab.~\ref{tab:qm_res_A}. The models with a latent space corresponding to the restricted output for the quantum layer are indicated with a subscript $1$ in the table. 

\begin{table}[]
\centering
\resizebox{\columnwidth}{!}{
\begin{tabular}{l|l|l|l|l|l|l}
model & $M_0$ & $M_1$ & $K_{tot}$ & Accuracy& TR$_{90}$ & VR$_{90}$ \\ \hline
 $qA(1)$ & 3 & 2 & 2358 & $0.9837^{+30}_{-36}$ & 0.9898& 0.9872  \\  \hline
$qB(2)$  & 6 & 4 & 4715 & $0.9847^{+32}_{-32}$  & 0.9920& 0.9889 \\
$qB_1(2)$  & 6 & 1 & 4713 & $0.9854^{+16}_{-20}$ & 0.9921 & 0.9887 \\ \hline
$qC(4)$ & 12 & 16 & 9437 & $0.9859^{+26}_{-18}$  & 0.9948 & 0.9891\\
$qC_1(4)$ & 12 & 1 & 9423 & $0.9839^{+26}_{-16}$ & 0.9905 & 0.9868 \\ \hline
$qD(6)$ & 18 & 64 & 14195 & $0.9869^{+24}_{-38}$ & 0.9948 & 0.9904 \\
$qD_1(6)$  & 18 & 1 & 14133 & $0.98125^{+38}_{-34}$ & 0.9881& 0.9858
\end{tabular}
}
\caption{Results for the first set of separable quantum models described in the text. The parenthesis in the label for the quantum models indicates the number of qubits $N$ employed.\label{tab:qm_res_A}}
\end{table}

\begin{figure}[]
	\centering
	 \includegraphics[width=0.45\textwidth]{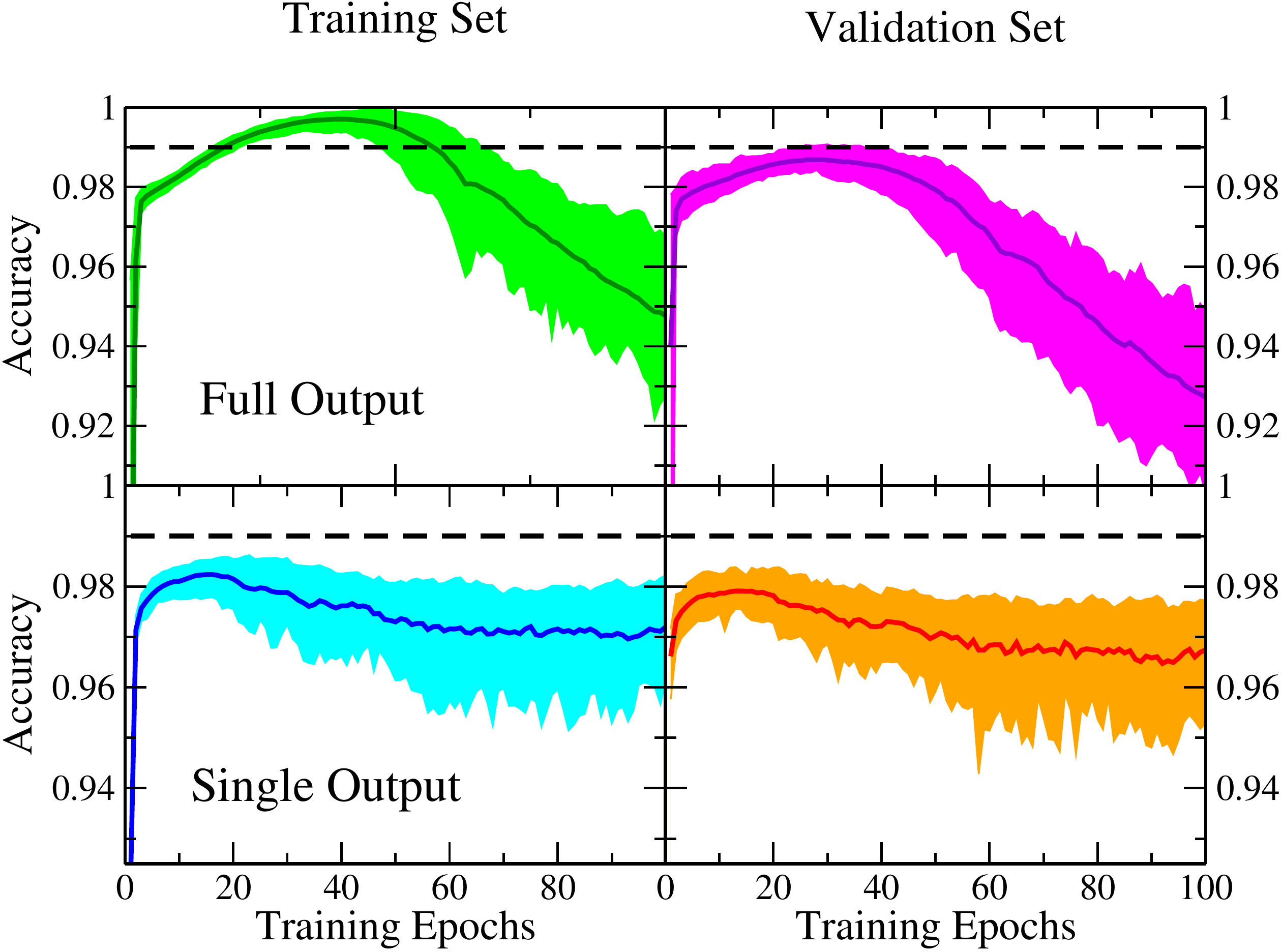}
	\caption{ Example of training of the largest quantum model described in the main text. The left panels show the increase in classification accuracy for the training set as a function of the number of epochs. The right panels shows the same for the validation set. The top panels are for the full model with the maximum number of outputs $M_1=16$ while the bottom panels are for the minimal model with a single output. The dashed black line indicates $99\%$ accuracy. All bands are $90\%$ confidence intervals with means indicated by the solid lines.}
	\label{fig:qm_train}
\end{figure}

The same convention is used in Tab.~\ref{tab:qm_res_B} where we present the parameters and results for the quantum models with entanglement described in Sec.~\ref{sec:qm_res_withent}. We also show in Fig.~\ref{fig:qm_train} the evolution of the accuracy for both the training set (left panels) and validation set (right panels). The top two panels are obtained with models with maximal output size ($M_1=16$ in this case) while the bottom panels show the results using a restricted output model with $M_1=1$.

\begin{table}[]
\centering
\resizebox{\columnwidth}{!}{
\begin{tabular}{l|l|l|l|l|l|l}
model  & $M_0$ & $M_1$ & $K_{tot}$ & Accuracy& TR$_{90}$ & VR$_{90}$ \\ \hline
$qE(2)$  & 9 & 4 & 7070 & $0.9879^{+12}_{-24}$ & 0.9940 & 0.9899 \\
$qE_1(2)$ & 9 & 1 & 7068 & $0.9865^{+22}_{-18}$ & 0.9933 & 0.9892 \\ \hline
$qF(2)$  & 15 & 4 & 11780 & $0.9875^{+18}_{-24}$ & 0.9960 & 0.99 \\
$qF_1(2)$  & 15 & 1 & 11778 & $0.9859^{+24}_{-16}$ & 0.9926 & 0.9893 \\ \hline
$qG(4)$ & 30 & 16 & 23567 & $0.9889^{+28}_{-22}$ & 0.9987 & 0.9924 \\
$qG_1(4)$ & 30 & 1 & 23553 & $0.9847^{+24}_{-34}$ & 0.9897 & 0.9879 \\ \hline
$qH(4)$ & 39 & 16 & 30632 & $0.9885^{+24}_{-22}$ & 0.9984 & 0.9915 \\
$qH_1(4)$& 39 & 1 & 30618 & $0.9833^{+18}_{-30}$ & 0.9867 & 0.9854 \\ \hline
$qI(4)$ &  57 & 16 & 44762 & $0.9883^{+22}_{-24}$ & 0.9993 & 0.9911 \\
$qI_1(4)$  & 57 & 1 & 44748 & $0.9810^{+24}_{-16}$ & 0.9863 & 0.9840 \\  \hline
$qL(6)$  & 45 & 64 & 35390 & $0.9895^{+18}_{-16}$ & 0.9995 & 0.9920 \\
$qL_1(6)$ & 45 & 1 & 35328 & $0.9823^{+32}_{-34}$ & 0.9871 & 0.9867 \\ \hline
\end{tabular}}
\caption{Results for the  set of quantum models described in the text. The parenthesis in the label for the quantum models indicates the number of qubits $N$ employed.\label{tab:qm_res_B}}
\end{table}

\begin{figure}[]
	\centering
	 \includegraphics[width=0.45\textwidth]{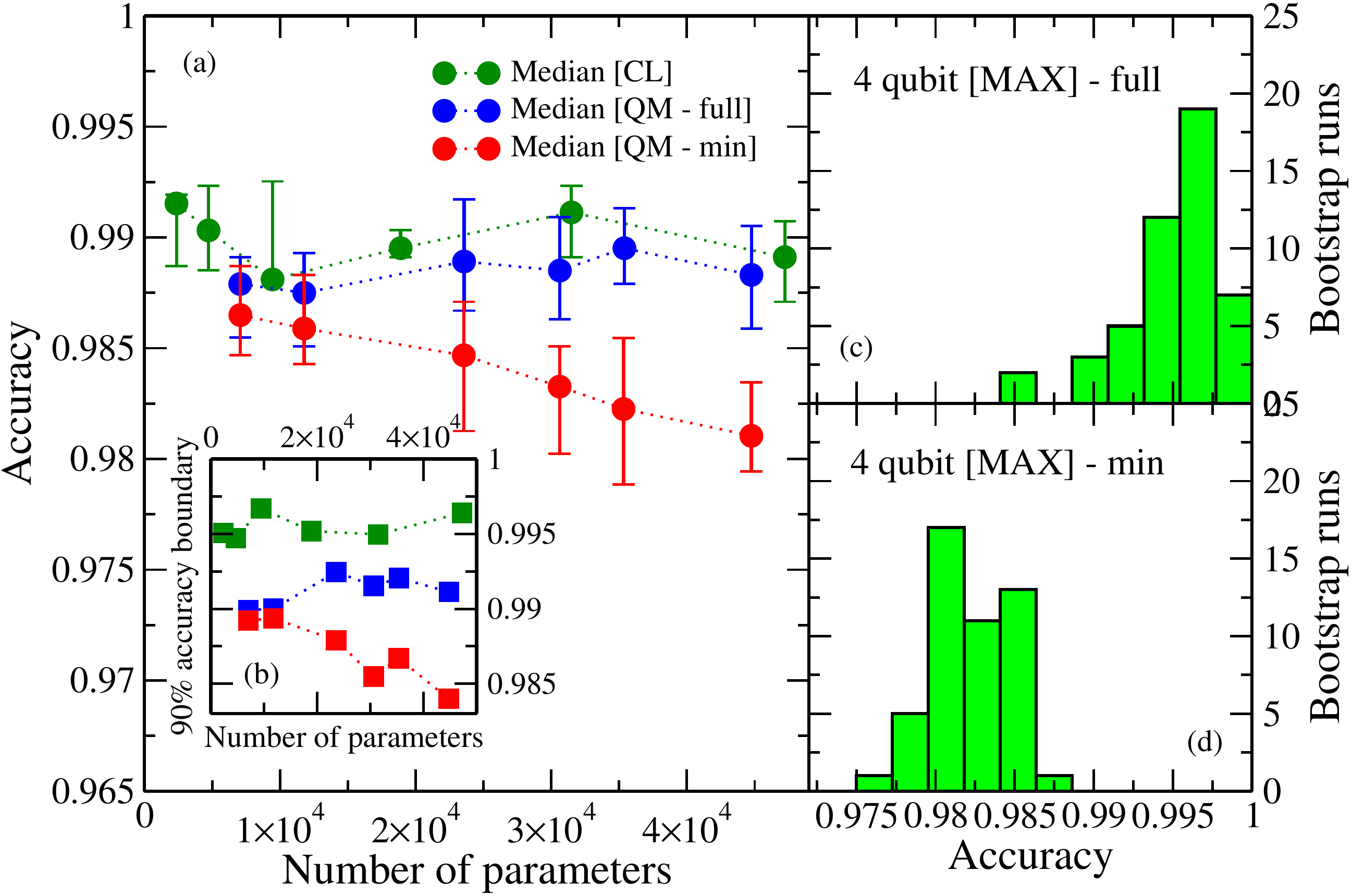}
	\caption{ Results for the accuracy achieved on MNIST with the classical model (green points) and the quantum models with entanglement described in the text (red and blue points). Panel (a) shows the accuracy as a function of the number of parameters for the classical model (green solid circles), the full quantum models (blue solid circles) and the quantum models with a single output variable $M_1=1$ (red solid circles) indicated by ``min''. The inset panel (b) shows the location of the $90\%$ accuracy percentile for the classical (green squares), full quantum (blue squares) and minimal quantum model (red squares) with a single output. Panels (c) and (d) show the histogram of achieved accuracies for the largest quantum model considered using $N=4$ qubits with either the full number of possible output variables ($M_1=16$, top panel) and the single output (bottom panel). }
	\label{fig:aqm_accuracy}
\end{figure}

Finally, we present in Fig.~\ref{fig:aqm_accuracy} an extension of the results presented in Fig.~\ref{fig:qm_accuracy_bounds} where in the left panels we show the accuracy histograms for the largest model considered in this work for both the full output model (panel(c)) and the restricted output model (panel(d)).

\begin{thebibliography}{39}
\providecommand{\natexlab}[1]{#1}
\providecommand{\url}[1]{\texttt{#1}}
\expandafter\ifx\csname urlstyle\endcsname\relax
  \providecommand{\doi}[1]{doi: #1}\else
  \providecommand{\doi}{doi: \begingroup \urlstyle{rm}\Url}\fi

\bibitem[Dunjko and Wittek(2020)]{Dunjko2020}
Vedran Dunjko and Peter Wittek.
\newblock A non-review of {Q}uantum {M}achine {L}earning: trends and
  explorations.
\newblock \emph{{Quantum Views}}, 4:\penalty0 32, March 2020.
\newblock \doi{10.22331/qv-2020-03-17-32}.
\newblock URL \url{https://doi.org/10.22331/qv-2020-03-17-32}.

\bibitem[Wiebe(2020)]{Wiebe2020}
Nathan Wiebe.
\newblock Key questions for the quantum machine learner to ask themselves.
\newblock \emph{New Journal of Physics}, 22\penalty0 (9):\penalty0 091001, sep
  2020.
\newblock \doi{10.1088/1367-2630/abac39}.
\newblock URL \url{https://doi.org/10.1088/1367-2630/abac39}.

\bibitem[Guan et~al.(2021)Guan, Perdue, Pesah, Schuld, Terashi, Vallecorsa, and
  Vlimant]{Guan_2021}
Wen Guan, Gabriel Perdue, Arthur Pesah, Maria Schuld, Koji Terashi, Sofia
  Vallecorsa, and Jean-Roch Vlimant.
\newblock Quantum machine learning in high energy physics.
\newblock \emph{Machine Learning: Science and Technology}, 2\penalty0
  (1):\penalty0 011003, Mar 2021.
\newblock ISSN 2632-2153.
\newblock \doi{10.1088/2632-2153/abc17d}.
\newblock URL \url{http://dx.doi.org/10.1088/2632-2153/abc17d}.

\bibitem[Cerezo et~al.(2021)Cerezo, Arrasmith, Babbush, Benjamin, Endo, Fujii,
  McClean, Mitarai, Yuan, Cincio, and Coles]{Cerezo_2021}
M.~Cerezo, Andrew Arrasmith, Ryan Babbush, Simon~C. Benjamin, Suguru Endo,
  Keisuke Fujii, Jarrod~R. McClean, Kosuke Mitarai, Xiao Yuan, Lukasz Cincio,
  and Patrick~J. Coles.
\newblock Variational quantum algorithms.
\newblock \emph{Nature Reviews Physics}, 3\penalty0 (9):\penalty0 625--644, aug
  2021.
\newblock \doi{10.1038/s42254-021-00348-9}.
\newblock URL \url{https://doi.org/10.1038/s42254-021-00348-9}.

\bibitem[Rebentrost et~al.(2014)Rebentrost, Mohseni, and Lloyd]{Rebentrost2014}
Patrick Rebentrost, Masoud Mohseni, and Seth Lloyd.
\newblock Quantum support vector machine for big data classification.
\newblock \emph{Phys. Rev. Lett.}, 113:\penalty0 130503, Sep 2014.
\newblock \doi{10.1103/PhysRevLett.113.130503}.
\newblock URL \url{https://link.aps.org/doi/10.1103/PhysRevLett.113.130503}.

\bibitem[Biamonte et~al.(2017)Biamonte, Wittek, Pancotti, Rebentrost, Wiebe,
  and Lloyd]{Biamonte2017}
Jacob Biamonte, Peter Wittek, Nicola Pancotti, Patrick Rebentrost, Nathan
  Wiebe, and Seth Lloyd.
\newblock Quantum machine learning.
\newblock \emph{Nature}, 549:\penalty0 195--202, 2017.
\newblock \doi{10.1038/nature23474}.
\newblock URL \url{https://doi.org/10.1038/nature23474}.

\bibitem[Huang et~al.(2021{\natexlab{a}})Huang, Broughton, Mohseni, Babbush,
  Boixo, Neven, and McClean]{Huang_2021}
Hsin-Yuan Huang, Michael Broughton, Masoud Mohseni, Ryan Babbush, Sergio Boixo,
  Hartmut Neven, and Jarrod~R. McClean.
\newblock Power of data in quantum machine learning.
\newblock \emph{Nature Communications}, 12\penalty0 (1), may
  2021{\natexlab{a}}.
\newblock \doi{10.1038/s41467-021-22539-9}.
\newblock URL \url{https://doi.org/10.1038/s41467-021-22539-9}.

\bibitem[Huang et~al.(2021{\natexlab{b}})Huang, Kueng, and
  Preskill]{Huang_2021B}
Hsin-Yuan Huang, Richard Kueng, and John Preskill.
\newblock Information-theoretic bounds on quantum advantage in machine
  learning.
\newblock \emph{Phys. Rev. Lett.}, 126:\penalty0 190505, May
  2021{\natexlab{b}}.
\newblock \doi{10.1103/PhysRevLett.126.190505}.
\newblock URL \url{https://link.aps.org/doi/10.1103/PhysRevLett.126.190505}.

\bibitem[Liu et~al.(2021)Liu, Arunachalam, and Temme]{Liu_2021}
Yunchao Liu, Srinivasan Arunachalam, and Kristan Temme.
\newblock A rigorous and robust quantum speed-up in supervised machine
  learning.
\newblock \emph{Nature Physics}, 17\penalty0 (9):\penalty0 1013--1017, jul
  2021.
\newblock \doi{10.1038/s41567-021-01287-z}.
\newblock URL \url{https://doi.org/10.1038/s41567-021-01287-z}.

\bibitem[Schuld and Killoran(2019)]{Schuld_2019b}
Maria Schuld and Nathan Killoran.
\newblock Quantum machine learning in feature hilbert spaces.
\newblock \emph{Phys. Rev. Lett.}, 122:\penalty0 040504, Feb 2019.
\newblock \doi{10.1103/PhysRevLett.122.040504}.
\newblock URL \url{https://link.aps.org/doi/10.1103/PhysRevLett.122.040504}.

\bibitem[McClean et~al.(2018)McClean, Boixo, Smelyanskiy, Babbush, and
  Neven]{McClean_2018}
Jarrod~R. McClean, Sergio Boixo, Vadim~N. Smelyanskiy, Ryan Babbush, and
  Hartmut Neven.
\newblock Barren plateaus in quantum neural network training landscapes.
\newblock \emph{Nature Communications}, 9\penalty0 (1), nov 2018.
\newblock \doi{10.1038/s41467-018-07090-4}.
\newblock URL \url{https://doi.org/10.1038/s41467-018-07090-4}.

\bibitem[Beer et~al.(2020)Beer, Bondarenko, Farrelly, Osborne, Salzmann,
  Scheiermann, and Wolf]{Beer_2020}
Kerstin Beer, Dmytro Bondarenko, Terry Farrelly, Tobias~J. Osborne, Robert
  Salzmann, Daniel Scheiermann, and Ramona Wolf.
\newblock Training deep quantum neural networks.
\newblock \emph{Nature Communications}, 11\penalty0 (1), feb 2020.
\newblock \doi{10.1038/s41467-020-14454-2}.
\newblock URL \url{https://doi.org/10.1038/s41467-020-14454-2}.

\bibitem[Ortiz~Marrero et~al.(2021)Ortiz~Marrero, Kieferov\'a, and
  Wiebe]{marrero2021entanglement}
Carlos Ortiz~Marrero, M\'aria Kieferov\'a, and Nathan Wiebe.
\newblock Entanglement-induced barren plateaus.
\newblock \emph{PRX Quantum}, 2:\penalty0 040316, Oct 2021.
\newblock \doi{10.1103/PRXQuantum.2.040316}.
\newblock URL \url{https://link.aps.org/doi/10.1103/PRXQuantum.2.040316}.

\bibitem[Pesah et~al.(2021)Pesah, Cerezo, Wang, Volkoff, Sornborger, and
  Coles]{Pesah2021}
Arthur Pesah, M.~Cerezo, Samson Wang, Tyler Volkoff, Andrew~T. Sornborger, and
  Patrick~J. Coles.
\newblock Absence of barren plateaus in quantum convolutional neural networks.
\newblock \emph{Phys. Rev. X}, 11:\penalty0 041011, Oct 2021.
\newblock \doi{10.1103/PhysRevX.11.041011}.
\newblock URL \url{https://link.aps.org/doi/10.1103/PhysRevX.11.041011}.

\bibitem[Banchi et~al.(2021)Banchi, Pereira, and Pirandola]{Banchi2021}
Leonardo Banchi, Jason Pereira, and Stefano Pirandola.
\newblock Generalization in quantum machine learning: A quantum information
  standpoint.
\newblock \emph{PRX Quantum}, 2:\penalty0 040321, Nov 2021.
\newblock \doi{10.1103/PRXQuantum.2.040321}.
\newblock URL \url{https://link.aps.org/doi/10.1103/PRXQuantum.2.040321}.

\bibitem[Du et~al.(2022)Du, Tu, Yuan, and Tao]{Du2022}
Yuxuan Du, Zhuozhuo Tu, Xiao Yuan, and Dacheng Tao.
\newblock Efficient measure for the expressivity of variational quantum
  algorithms.
\newblock \emph{Phys. Rev. Lett.}, 128:\penalty0 080506, Feb 2022.
\newblock \doi{10.1103/PhysRevLett.128.080506}.
\newblock URL \url{https://link.aps.org/doi/10.1103/PhysRevLett.128.080506}.

\bibitem[Havlíček et~al.(2019)Havlíček, Córcoles, Temme, Harrow, Kandala,
  Chow, and Gambetta]{Havl_ek_2019}
Vojtěch Havlíček, Antonio~D. Córcoles, Kristan Temme, Aram~W. Harrow,
  Abhinav Kandala, Jerry~M. Chow, and Jay~M. Gambetta.
\newblock Supervised learning with quantum-enhanced feature spaces.
\newblock \emph{Nature}, 567\penalty0 (7747):\penalty0 209–212, Mar 2019.
\newblock ISSN 1476-4687.
\newblock \doi{10.1038/s41586-019-0980-2}.
\newblock URL \url{http://dx.doi.org/10.1038/s41586-019-0980-2}.

\bibitem[Peters et~al.(2021)Peters, Caldeira, Ho, Leichenauer, Mohseni, Neven,
  Spentzouris, Strain, and Perdue]{peters2021machine}
Evan Peters, João Caldeira, Alan Ho, Stefan Leichenauer, Masoud Mohseni,
  Hartmut Neven, Panagiotis Spentzouris, Doug Strain, and Gabriel~N. Perdue.
\newblock Machine learning of high dimensional data on a noisy quantum
  processor.
\newblock \emph{npj Quantum Information}, \penalty0 (7):\penalty0 161, 2021.
\newblock \doi{10.1038/s41534-021-00498-9}.
\newblock URL \url{https://doi.org/10.1038/s41534-021-00498-9}.

\bibitem[Abadi et~al.(2015)Abadi, Agarwal, Barham, Brevdo, Chen, Citro,
  Corrado, Davis, Dean, Devin, Ghemawat, Goodfellow, Harp, Irving, Isard, Jia,
  Jozefowicz, Kaiser, Kudlur, Levenberg, Man\'{e}, Monga, Moore, Murray, Olah,
  Schuster, Shlens, Steiner, Sutskever, Talwar, Tucker, Vanhoucke, Vasudevan,
  Vi\'{e}gas, Vinyals, Warden, Wattenberg, Wicke, Yu, and
  Zheng]{tensorflow2015-whitepaper}
Mart\'{\i}n Abadi, Ashish Agarwal, Paul Barham, Eugene Brevdo, Zhifeng Chen,
  Craig Citro, Greg~S. Corrado, Andy Davis, Jeffrey Dean, Matthieu Devin,
  Sanjay Ghemawat, Ian Goodfellow, Andrew Harp, Geoffrey Irving, Michael Isard,
  Yangqing Jia, Rafal Jozefowicz, Lukasz Kaiser, Manjunath Kudlur, Josh
  Levenberg, Dandelion Man\'{e}, Rajat Monga, Sherry Moore, Derek Murray, Chris
  Olah, Mike Schuster, Jonathon Shlens, Benoit Steiner, Ilya Sutskever, Kunal
  Talwar, Paul Tucker, Vincent Vanhoucke, Vijay Vasudevan, Fernanda Vi\'{e}gas,
  Oriol Vinyals, Pete Warden, Martin Wattenberg, Martin Wicke, Yuan Yu, and
  Xiaoqiang Zheng.
\newblock {TensorFlow}: Large-scale machine learning on heterogeneous systems,
  2015.
\newblock URL \url{https://www.tensorflow.org/}.
\newblock Software available from tensorflow.org.

\bibitem[Paszke et~al.(2019)Paszke, Gross, Massa, Lerer, Bradbury, Chanan,
  et~al.]{PyTorch}
Adam Paszke, Sam Gross, Francisco Massa, Adam Lerer, James Bradbury, Gregory
  Chanan, et~al.
\newblock Pytorch: An imperative style, high-performance deep learning library.
\newblock In H.~Wallach, H.~Larochelle, A.~Beygelzimer, F.~d\textquotesingle
  Alch\'{e}-Buc, E.~Fox, and R.~Garnett, editors, \emph{Advances in Neural
  Information Processing Systems 32}, pages 8026--8037. Curran Associates,
  Inc., 2019.
\newblock URL
  \url{http://papers.nips.cc/paper/9015-pytorch-an-imperative-style-high-performance-deep-learning-library.pdf}.

\bibitem[Filipek et~al.(2021)Filipek, Roggero, Hsu, and Wiebe]{squid}
J.~Filipek, A.~Roggero, S.~Hsu, and N.~Wiebe.
\newblock {SQUID}.
\newblock \url{https://bitbucket.org/squid-qml/}, 2021.

\bibitem[Smith et~al.(2016)Smith, Curtis, and Zeng]{smith2016practical}
Robert~S. Smith, Michael~J. Curtis, and William~J. Zeng.
\newblock A practical quantum instruction set architecture.
\newblock \emph{arXiv}, \penalty0 (1608.03355), 2016.
\newblock \doi{10.48550/ARXIV.1608.03355}.

\bibitem[et~al.(2019)]{Qiskit}
H{\'e}ctor~Abraham et~al.
\newblock Qiskit: An open-source framework for quantum computing.
\newblock 2019.
\newblock \doi{10.5281/zenodo.2562110}.

\bibitem[team and collaborators(2020)]{Cirq}
Quantum~AI team and collaborators.
\newblock Cirq.
\newblock October 2020.
\newblock \doi{10.5281/zenodo.4062499}.
\newblock URL \url{https://doi.org/10.5281/zenodo.4062499}.

\bibitem[Harris et~al.(2020)Harris, Millman, van~der Walt, Gommers, Virtanen,
  Cournapeau, et~al.]{numpy}
Charles~R. Harris, K.~Jarrod Millman, St{'{e}}fan~J. van~der Walt, Ralf
  Gommers, Pauli Virtanen, David Cournapeau, et~al.
\newblock Array programming with {NumPy}.
\newblock \emph{Nature}, 585\penalty0 (7825):\penalty0 357--362, September
  2020.
\newblock \doi{10.1038/s41586-020-2649-2}.
\newblock URL \url{https://doi.org/10.1038/s41586-020-2649-2}.

\bibitem[Brassard et~al.(2002)Brassard, Høyer, Mosca, and Tapp]{Brassard_2002}
Gilles Brassard, Peter Høyer, Michele Mosca, and Alain Tapp.
\newblock Quantum amplitude amplification and estimation.
\newblock \emph{Quantum Computation and Information}, page 53–74, 2002.
\newblock ISSN 0271-4132.
\newblock \doi{10.1090/conm/305/05215}.
\newblock URL \url{http://dx.doi.org/10.1090/conm/305/05215}.

\bibitem[Suzuki et~al.(2020)Suzuki, Uno, Raymond, Tanaka, Onodera, and
  Yamamoto]{Suzuki_2020}
Yohichi Suzuki, Shumpei Uno, Rudy Raymond, Tomoki Tanaka, Tamiya Onodera, and
  Naoki Yamamoto.
\newblock Amplitude estimation without phase estimation.
\newblock \emph{Quantum Information Processing}, 19\penalty0 (2), Jan 2020.
\newblock ISSN 1573-1332.
\newblock \doi{10.1007/s11128-019-2565-2}.
\newblock URL \url{http://dx.doi.org/10.1007/s11128-019-2565-2}.

\bibitem[Grinko et~al.(2021)Grinko, Gacon, Zoufal, and Woerner]{Grinko_2021}
Dmitry Grinko, Julien Gacon, Christa Zoufal, and Stefan Woerner.
\newblock Iterative quantum amplitude estimation.
\newblock \emph{npj Quantum Information}, 7\penalty0 (1), Mar 2021.
\newblock ISSN 2056-6387.
\newblock \doi{10.1038/s41534-021-00379-1}.
\newblock URL \url{http://dx.doi.org/10.1038/s41534-021-00379-1}.

\bibitem[Gily{\'e}n et~al.(2019)Gily{\'e}n, Arunachalam, and
  Wiebe]{gilyen2019optimizing}
Andr{\'a}s Gily{\'e}n, Srinivasan Arunachalam, and Nathan Wiebe.
\newblock Optimizing quantum optimization algorithms via faster quantum
  gradient computation.
\newblock In \emph{Proceedings of the Thirtieth Annual ACM-SIAM Symposium on
  Discrete Algorithms}, pages 1425--1444. SIAM, 2019.
\newblock \doi{10.1137/1.9781611975482.87}.

\bibitem[Schuld et~al.(2020)Schuld, Bocharov, Svore, and Wiebe]{Schuld2020}
Maria Schuld, Alex Bocharov, Krysta~M. Svore, and Nathan Wiebe.
\newblock Circuit-centric quantum classifiers.
\newblock \emph{Phys. Rev. A}, 101:\penalty0 032308, Mar 2020.
\newblock \doi{10.1103/PhysRevA.101.032308}.
\newblock URL \url{https://link.aps.org/doi/10.1103/PhysRevA.101.032308}.

\bibitem[Vidal and Dawson(2004)]{vidal2004}
G.~Vidal and C.~M. Dawson.
\newblock Universal quantum circuit for two-qubit transformations with three
  controlled-not gates.
\newblock \emph{Phys. Rev. A}, 69:\penalty0 010301, Jan 2004.
\newblock \doi{10.1103/PhysRevA.69.010301}.
\newblock URL \url{https://link.aps.org/doi/10.1103/PhysRevA.69.010301}.

\bibitem[Vatan and Williams(2004)]{vatan2004}
Farrokh Vatan and Colin Williams.
\newblock Optimal quantum circuits for general two-qubit gates.
\newblock \emph{Phys. Rev. A}, 69:\penalty0 032315, Mar 2004.
\newblock \doi{10.1103/PhysRevA.69.032315}.
\newblock URL \url{https://link.aps.org/doi/10.1103/PhysRevA.69.032315}.

\bibitem[{Lecun} et~al.(1998){Lecun}, {Bottou}, {Bengio}, and {Haffner}]{MNIST}
Y.~{Lecun}, L.~{Bottou}, Y.~{Bengio}, and P.~{Haffner}.
\newblock Gradient-based learning applied to document recognition.
\newblock \emph{Proceedings of the IEEE}, 86\penalty0 (11):\penalty0
  2278--2324, 1998.
\newblock \doi{10.1109/5.726791}.

\bibitem[Bergholm et~al.(2018)Bergholm, Izaac, Schuld, Gogolin, Alam, Ahmed,
  Arrazola, Blank, Delgado, Jahangiri, McKiernan, Meyer, Niu, Száva, and
  Killoran]{Bergholm2020Pennylane}
Ville Bergholm, Josh Izaac, Maria Schuld, Christian Gogolin, M.~Sohaib Alam,
  Shahnawaz Ahmed, Juan~Miguel Arrazola, Carsten Blank, Alain Delgado, Soran
  Jahangiri, Keri McKiernan, Johannes~Jakob Meyer, Zeyue Niu, Antal Száva, and
  Nathan Killoran.
\newblock Pennylane: Automatic differentiation of hybrid quantum-classical
  computations.
\newblock \emph{arXiv}, \penalty0 (1811.04968), 2018.
\newblock \doi{10.48550/ARXIV.1811.04968}.
\newblock URL \url{https://arxiv.org/abs/1811.04968}.

\bibitem[Shen et~al.(2020)Shen, Zhang, You, and Zhai]{Shen2020}
Huitao Shen, Pengfei Zhang, Yi-Zhuang You, and Hui Zhai.
\newblock Information scrambling in quantum neural networks.
\newblock \emph{Phys. Rev. Lett.}, 124:\penalty0 200504, May 2020.
\newblock \doi{10.1103/PhysRevLett.124.200504}.
\newblock URL \url{https://link.aps.org/doi/10.1103/PhysRevLett.124.200504}.

\bibitem[Liu et~al.(2019)Liu, Ran, Wittek, Peng, Garc{\'{\i}}a, Su, and
  Lewenstein]{Liu_2019}
Ding Liu, Shi-Ju Ran, Peter Wittek, Cheng Peng, Raul~Bl{\'{a}}zquez
  Garc{\'{\i}}a, Gang Su, and Maciej Lewenstein.
\newblock Machine learning by unitary tensor network of hierarchical tree
  structure.
\newblock \emph{New Journal of Physics}, 21\penalty0 (7):\penalty0 073059, jul
  2019.
\newblock \doi{10.1088/1367-2630/ab31ef}.
\newblock URL \url{https://doi.org/10.1088/1367-2630/ab31ef}.

\bibitem[Roberts et~al.(2019)Roberts, Milsted, Ganahl, Zalcman, Fontaine, Zou,
  Hidary, Vidal, and Leichenauer]{roberts2019tensornetwork}
Chase Roberts, Ashley Milsted, Martin Ganahl, Adam Zalcman, Bruce Fontaine,
  Yijian Zou, Jack Hidary, Guifre Vidal, and Stefan Leichenauer.
\newblock Tensornetwork: A library for physics and machine learning.
\newblock \emph{arXiv}, \penalty0 (1905.01330), 2019.
\newblock \doi{10.48550/ARXIV.1905.01330}.

\bibitem[Mitarai et~al.(2018)Mitarai, Negoro, Kitagawa, and
  Fujii]{Mitarai_2018}
K.~Mitarai, M.~Negoro, M.~Kitagawa, and K.~Fujii.
\newblock Quantum circuit learning.
\newblock \emph{Physical Review A}, 98\penalty0 (3), Sep 2018.
\newblock ISSN 2469-9934.
\newblock \doi{10.1103/physreva.98.032309}.
\newblock URL \url{http://dx.doi.org/10.1103/PhysRevA.98.032309}.

\bibitem[Schuld et~al.(2019)Schuld, Bergholm, Gogolin, Izaac, and
  Killoran]{Schuld_2019}
Maria Schuld, Ville Bergholm, Christian Gogolin, Josh Izaac, and Nathan
  Killoran.
\newblock Evaluating analytic gradients on quantum hardware.
\newblock \emph{Physical Review A}, 99\penalty0 (3), Mar 2019.
\newblock ISSN 2469-9934.
\newblock \doi{10.1103/physreva.99.032331}.
\newblock URL \url{http://dx.doi.org/10.1103/PhysRevA.99.032331}.

\end{thebibliography}
\end{document}